\documentclass[aps,preprint,eqsecnum]{revtex4}

\usepackage{bm}
\usepackage{amsmath}
\usepackage{amssymb}
\usepackage{amsfonts}
\usepackage{graphicx,color}
\usepackage{isorot}

\usepackage{graphicx,color}
\usepackage{float}
\restylefloat{figure}
\newcommand{\bj}{\boldsymbol{j}}

\begin{document}

\title[Molecules in fields]
{Classical mechanics of dipolar asymmetric top molecules in
collinear static electric and nonresonant linearly polarized laser fields:
energy-momentum diagrams, bifurcations and accessible configuration space}

\author{Carlos A. Arango}
\altaffiliation[Current address: ]{Department of Chemistry, University of Toronto,
Toronto, Ontario M5S 3H6, Canada}
\author{Gregory S. Ezra}
\email{gse1@cornell.edu}
\affiliation{Department of Chemistry and Chemical Biology\\
Baker Laboratory\\
Cornell University\\
Ithaca\\
NY 14853\\
}

\date{\today}

\begin{abstract}

We study classical energy-momentum ($E$-$m$) diagrams for rotational motion of
dipolar asymmetric top molecules in strong external fields.  Static electric fields,
nonresonant linearly polarized laser fields, and collinear combinations of the two
are investigated.
We treat specifically the molecules iodobenzene (a nearly prolate asymmetric top), pyridazine
(nearly oblate asymmetric top), and iodopentafluorobenzene (intermediate case).
The location of relative equilibria in the $E$-$m$ plane and associated bifurcations
are determined by straightforward calculation, with
analytical results given where possible. In cases where
analytical solutions cannot be obtained,
we resort to numerical solutions, while keeping a geometrical picture of the
nature of the solutions to the fore.
The classification we obtain of the topology of classically allowed rotor configuration space
regions in the $E$-$m$ diagram
is of potential use in characterization of
energy eigenstates of the corresponding quantum mechanical problem.

\end{abstract}



\maketitle

\newpage

\section{Introduction}
\label{Introduction}

The study of rigid body motion is one of the most important topics in classical and quantum mechanics
\cite{Klein65,Casimir31,Deprit67,Arnold78,Zare88,Arnold88,Deprit93,Goldstein02}.
Of particular importance are the integrable cases of the rigid body problem \cite{Oshemkov91,Bolsinov04},
which include the free asymmetric top (Euler top) \cite{Klein65,Arnold78,Arnold88},
the symmetric top in a uniform external gravitational field (Lagrange top) \cite{Arnold88,Cushman97},
and the Kovalevskaya top \cite{Arnold88,Perelomov02}.
The theory of rigid body motion also provides the basis for analysis and interpretation of
the rotational dynamics and spectra of semi-rigid molecules
\cite{Casimir31,Townes75,Zare88,Harter88,Kroto92,Bunker98}.
In molecular terms the Euler top is simply a free asymmetric top molecule \cite{Zare88,Harter88,Kroto92},
the Lagrange top models a symmetric top molecule with a dipole moment in an electric field
\cite{Kroto92,Kozin03}, while there does not appear to be an obvious molecular analogue
for the Kovalevskaya top.
Breaking the symmetry of the moment of inertia tensor in the Lagrange top results in
the non-integrable problem of an asymmetric top in a static field.
Molecular examples of these two cases are iodobenzene (near prolate) \cite{Peronne03,Peronne04} and
pyridazine (near oblate) \cite{Li98} in a static electric field.
The rotational constants of the molecule  iodopentafluorobenzene make it a more  generic example of an
asymmetric rotor \cite{Poulsen04}.

The general problem of classical-quantum correspondence \cite{Gutzwiller90,Child91}
is of great interest for both integrable and nonintegrable rotor systems.
Various aspects of the classical-quantum correspondence have been studied for
diatomic molecules in tilted fields, {\it{i.e.}},
noncollinear static electric  and nonresonant linearly polarized laser fields \cite{Arango05}.
The integrable collinear case exhibits the phenomenon of monodromy \cite{Cushman97}
both classically
and quantum mechanically \cite{Arango04}.
For the nonintegrable case of tilted fields the rotor motion tends to be
integrable in both low-energy (pendular) and high-energy (free-rotor) limits,
and chaotic at intermediate energies, with
the degree of chaos controllable by variation of the angle between the fields \cite{Arango05}.
For collinear fields the system is integrable,
with both the energy $E$ and the projection of the angular momentum into the space fixed $z$-axis, $m$,  as
constants of motion. The effective potential $V_{\text{eff}}(\theta; m)$ for a given value of $m$
exhibits extrema in the $\theta$ (polar) coordinate, which define the relative
equilibria \cite{Smale70,Arnold78,Arnold88}. Plotting the location of these extrema in the $E$-$m$ plane
gives the energy-momentum diagram for the system \cite{Arango05a}.
The $E$-$m$ diagram provides a useful global classification of the
rotor dynamics, as distinct regions of the $E$-$m$ plane are associated with
different allowed types of motion of the diatomic. For symmetric tops in
electric fields similar diagrams can be constructed \cite{Kozin03}, and analysis of classical symmetric top
$E$-$m$ and $E$-$k$ diagrams helps understand the organization of the quantum level spectrum \cite{Kozin03}.

The problem of a dipolar \emph{asymmetric} top in a static external field is nonintegrable
\cite{Grozdanov96,Arango05a}, as is the problem of a polarizable asymmetric top in a
nonresonant laser field \cite{Arango05a}.
For asymmetric tops in either static or laser fields or collinear superpositions
of the two, the angular momentum projection $m$ is a constant of motion.
Although the
complicated form of the kinetic energy does not allow us to
separate an effective potential as straightforwardly as in
the diatomic or the symmetric top case,
it is still possible to define an effective or amended potential \cite{Smale70,Arnold88,Bolsinov04}
for the class of motions in which
the asymmetric top molecule rotates with constant Euler angles $\theta$ and $\psi$
(the third Euler angle $\phi$ is an ignorable coordinate).
The energies of the extrema of this potential for
given values of $m$ again define an energy-momentum diagram
\cite{Smale70,Arnold78,Iacob71,Katok72,Tatarinov74,Artigue86},
which can be used to classify the motions of the asymmetric top.

The asymmetric top in an external static electric field is an example of a  dynamical system
with symmetry \cite{Smale70,Arnold88,Marsden92,Marsden99}.  The potential energy and Hamiltonian in
this case are invariant with respect to rotations about the space-fixed field direction.
There is therefore an associated constant of the motion $m$, the projection of the angular momentum
vector $\bj$ onto the external field axis, in addition to the energy $E$.
The mapping of the system phase space onto the $E$-$m$ plane is called the energy-momentum map
\cite{Smale70,Arnold88,Marsden92,Marsden99}.
Critical points of this mapping define the bifurcations sets in the $E$-$m$ plane, which
form the boundaries of regions of qualitatively different types of classical motion
\cite{Smale70,Bolsinov04}.
Following the fundamental work of Smale \cite{Smale70}, applications of these concepts
were made to integrable rotor problems \cite{Iacob71},
asymmetric rotors in an external gravitational field \cite{Katok72}, and symmetric \cite{Tatarinov73}
and asymmetric
\cite{Tatarinov74} rotors in more complicated potentials.
Relative equilibria have also been studied for rotating semi-rigid molecules in the absence of
external fields \cite{Montaldi99,Kozin00} and for transition states in rotationally inelastic collisions
\cite{Wiesenfeld03}.

In the present paper we study  $E$-$m$
diagrams of relative equilibria for molecular asymmetric tops for several molecule-field configurations of
physical interest \cite{Stapelfeldt03,Seideman05}: a static electric field; a
nonresonant linearly polarized laser field; both fields in a collinear combination.
For the field strengths considered, the free asymmetric top motion is strongly perturbed.
The associated $E$-$m$
diagrams are obtained for the most part analytically.  Our aim here is to understand
the nature of the rotor motions associated with
different regions and curves in the $E$-$m$ diagrams  for physically relevant values of the
field parameters.
We also study the topological
classification of the allowed $\theta$-$\psi$ configuration space of the system
in terms of their (multivalued) genus \cite{Arnold88}.
An extension of these results, to be discussed in a future paper, involves comparison of
computed quantum mechanical eigenstate probability densities with the boundaries of
classically allowed regions in  $\theta$-$\psi$ configuration space.
The comparisons indicate  that the classical mechanical
methods developed here provide a promising foundation for the difficult task of classifying
the quantum levels of the complex system consisting of an asymmetric rotor in
external fields \cite{Block92,Moore99,Kanya04}.

We mention that the approach adopted here
may be thought of as the analogue for perturbed rotor systems  of
the methods applied by Kellman and coworkers \cite{Kellman95,Rose00} to vibrational problems exhibiting a single
conserved vibrational (superpolyad) quantum number (see also Ref.\ \onlinecite{Cooper05}).

There have been many studies of the classical dynamics of a
rigid asymmetric top rotating about a fixed point
in a gravitational field (this is the classic heavy top problem); see, for example
\cite{Klein65,Iacob71,Katok72,Galgani81,Artigue86,Arnold88,ChavoyaAceves89,Lewis92,Broucke93,Gashenenko04}.
Katok \cite{Katok72} and Gashenenko and Richter \cite{Gashenenko04}
obtained $E$-$m$ diagrams and analyzed bifurcations of relative equilibria
in the $E$-$m$ plane.  These authors also classified the topology of accessible system configuration space.
We present here a similar analysis for several molecular examples of an asymmetric top molecule
possessing a dipole moment in a static electric field. In the spirit of Katok's analysis \cite{Katok72},
we use straightforward analytical and geometric methods to build
the $E$-$m$ diagram for the molecules of interest.

Katok's treatment was generalized by Tatarinov \cite{Tatarinov74} (see also \cite{Arnold88}) to include
more complicated gravitational perturbations of the rotational dynamics of an asymmetric top.
Although given for a specific potential, Tatarinov's analysis can be mapped directly
onto the molecular case of an asymmetric top in collinear fields \cite{Arango05a}.
Following Tatarinov, we obtain $E$-$m$ diagrams
for asymmetric molecules of physical interest, and study bifurcations and other
characteristic features of the problem, including the topology of the  $\theta$-$\psi$ configuration space and
its classification according to the genera of the (connected) allowed regions \cite{Arnold88}.

This paper is organized as follows. First in Section \ref{Hamiltonian} we derive
the Hamiltonian for asymmetric tops in the general tilted fields case, with the aim of
introducing our notation and conventions.
In Section \ref{em_electric} we treat the molecules in a static electric field,
while in Section \ref{em_laser} we study the same molecules in a nonresonant linearly
polarized laser field. In Section \ref{em_collinear} we analyse these molecules in collinear fields.
Section \ref{conclusions} concludes.

Finally we mention that, the advantages of alternative
approaches notwithstanding \cite{Cushman05},
all calculations reported here have been carried out using polar coordinates.

\newpage

\section{Hamiltonian for the Asymmetric Top in combined Fields}
\label{Hamiltonian}

\subsection{General case: tilted fields}

We use the $y$-convention  \cite{Zare88} to define
the three Euler angles $\left(\theta,\phi,\psi\right)$
describing the orientation of the body-fixed frame with respect to lab-fixed frame.
In the body-fixed frame the kinetic energy of the free asymmetric top can be written in
terms of the components of the angular momentum ${\bm j}=(j_1,j_2,j_3)$ and the three
components of the (diagonal) moment of inertia tensor
${\mathsf I}={\mathrm{diag}}(I_1,I_2,I_3)$
\begin{equation}\label{eqnnlr1}
T=\frac{j_1^2}{2I_1}+\frac{j_2^2}{2I_2}+\frac{j_3^2}{2I_3}.
\end{equation}
For an asymmetric top, $I_1 \neq I_2 \neq I_3$.
In terms of the Euler angles and their conjugate momenta
$\left(p_\theta,p_\phi,p_\psi\right)$
the body-fixed components of ${\bm j}$ are
\begin{subequations}\label{eqnnlr2}
\begin{align}
j_1&=p_\theta \sin\psi - p_\phi \frac{\cos\psi}{\sin\theta} + p_\psi \cos\psi \cot\theta,\\
j_2&=p_\theta \cos\psi + p_\phi \frac{\sin\psi}{\sin\theta} - p_\psi \sin\psi \cot\theta,\\
j_3&=p_\psi
\end{align}
\end{subequations}
so that
\begin{equation}\label{eqnnlr2.1}
\begin{split}
j^2&=j_1^2+j_2^2+j_3^2\\
&=p_\theta^2+\frac{1}{\sin^2\theta}\left(p_\phi-p_\psi \cos\theta\right)^2+p_\psi^2.
\end{split}
\end{equation}
From \eqref{eqnnlr1}, the kinetic energy $T=T(\theta,\phi,\psi,p_\theta,p_\phi,p_\psi)$ is then
\begin{equation}\label{eqnnlr3}
T=\frac{1}{2I_1}\left[p_\theta\sin\psi+\frac{\cos\psi}{\sin\theta}\left(p_\psi\cos\theta-p_\phi\right)\right]^2
+\frac{1}{2I_2}\left[p_\theta\cos\psi+\frac{\sin\psi}{\sin\theta}\left(p_\phi-p_\psi\cos\theta\right)\right]^2
+\frac{p_\psi^2}{2I_3}.
\end{equation}
Immediately we see that $\phi$ is an ignorable coordinate and $p_\phi=m$,
the projection of $\bm j$ into the space fixed $z$-axis,
is a constant of the motion for the free top.

If the polarizability in the molecule-fixed frame is given by the
diagonal tensor $\alpha = \text{diag}(\alpha_1, \alpha_2, \alpha_3)$,
the interaction with a nonresonant laser field polarized along the space-fixed $z$-axis is
\cite{Stapelfeldt03}
\begin{equation}\label{eqnnlr6.1}
V_L=-\frac{\varepsilon_L^2}{4}\left[\alpha_1+\left(\alpha_2-\alpha_1\right)\sin^2\theta \sin^2\psi
+ \left(\alpha_3-\alpha_1\right)\cos^2\theta \right],
\end{equation}
with $\varepsilon_L^2$ proportional to the intensity of the laser field.
Omitting the angle-independent term (which produces a constant shift in energy) we obtain
\begin{equation}\label{eqnnlr6.2}
V_L=-\Delta\omega_2\sin^2\theta \sin^2\psi - \Delta\omega_3\cos^2\theta,
\end{equation}
with $\Delta\omega_2=\left(\alpha_2-\alpha_1\right)\varepsilon_L^2/4$
and $\Delta\omega_3=\left(\alpha_3-\alpha_1\right)\varepsilon_L^2/4$.
Assuming the dipole moment to lie along the molecule-fixed ${\bar z}$-axis,
the interaction with a static electric field
tilted through an angle $\beta$ with respect to the space-fixed $z$-axis
and lying in the space-fixed $xy$-plane is
\begin{equation}\label{eqnnlr4}
V_S=-d_0 \varepsilon_S (\cos\beta \cos\theta + \sin\beta \cos\phi \sin\theta),
\end{equation}
where $d_0$ is the magnitude of the electric dipole moment and $\varepsilon_S$
is the strength of the static field.

The Hamiltonian of the asymmetric top in tilted fields can
be written from equations \eqref{eqnnlr3}, \eqref{eqnnlr4} and \eqref{eqnnlr6.2} as
\begin{equation}\label{eqnnlr7}
H=T+V_S+V_L.
\end{equation}
Note that the potential $V_S+V_L$ is a function of the angle $\phi$, so that
$m$ is not conserved.  The asymmetric top in tilted fields is therefore a
physically significant rotor problem with three degrees of freedom.
For collinear fields, $\beta=0$, the $\phi$ angle is not present in $H$ and $p_\phi=m$ is
then a constant of the motion.
In the remainder of this paper we consider the collinear case only.

\subsection{Collinear fields}

The Hamiltonian for an asymmetric top molecule in collinear fields can be rewritten as
\begin{equation}\label{eqnnlr63}
H=\tfrac{1}{2}({\mathsf I}^{-1}{\bm j})\cdot{\bm j}+V(\theta,\psi),
\end{equation}
with
\begin{equation}\label{eqnnlr63.1}
V(\theta,\psi)=-\omega\cos\theta-\Delta\omega_2\sin^2\theta \sin^2\psi - \Delta\omega_3\cos^2\theta.
\end{equation}
In terms of the Euler angles and their time derivatives the body-fixed
components of the angular momentum are
\begin{subequations}\label{eqnnlr64}
\begin{align}
j_1&=I_1\left(\dot\theta\sin\psi-\dot\phi\sin\theta\cos\psi\right),\label{eqnnlr64.1}\\
j_2&=I_2\left(\dot\theta\cos\psi+\dot\phi\sin\theta\sin\psi\right),\label{eqnnlr64.2}\\
j_3&=I_3\left(\dot\phi\cos\theta+\dot\psi\right).\label{eqnnlr64.3}
\end{align}
\end{subequations}
The projection of $\bm j$ onto the space-fixed unit vector ${\bm e}_z=(0,0,1)$
is given in terms of the direction cosine matrix ${\bm C}$ as
\begin{equation}\label{eqnnlr64.2a}
m 
=\tilde{\bm C}_3\cdot{\bm j},
\end{equation}
where
\begin{equation}\label{eqnnlr64.3a}
\begin{split}
\tilde{\bm C}_3&=({\tilde C}_{31},{\tilde C}_{32},{\tilde C}_{33})\\
&=(-\sin\theta\cos\psi,\sin\theta\sin\psi,\cos\theta),
\end{split}
\end{equation}
is the vector of body-fixed components of ${\bm e}_z$.

\subsection{Determination of relative equilibria}

As the angle $\phi$ is ignorable, we consider relative equilibria defined
by the conditions $\dot\theta=0$, and $\dot\psi=0$ \cite{Katok72}.
These relative equilibria in general define periodic orbits in the full rotor phase space.
Using these conditions in equations \eqref{eqnnlr64}, and rewriting in terms of $\tilde{\bm C}_3$ we obtain
\begin{equation}\label{eqnnlr65}
\bm j=\dot\phi({\mathsf I}\tilde{\bm C}_3).
\end{equation}
The equation for $m$, \eqref{eqnnlr64.2a}, can be rewritten as
\begin{equation}\label{eqnnlr66}
m=\dot\phi({\mathsf I}\tilde{\bm C}_3)\cdot\tilde{\bm C}_3,
\end{equation}
which can be used to express $\dot\phi$ in terms of $m$ and $\tilde{\bm C}_3$.
Substituting the resulting expression for $\bm j$,
\begin{equation}\label{eqnnlr67}
\bm j=\frac{m({\mathsf I}\tilde{\bm C}_3)}{({\mathsf I}\tilde{\bm C}_3)\cdot\tilde{\bm C}_3},
\end{equation}
into the Hamiltonian \eqref{eqnnlr63} gives the effective or amended potential
\cite{Katok72,Tatarinov74,Arnold88,Bolsinov04} for an asymmetric top molecule in collinear fields
\begin{equation}\label{eqnnlr68}
V_m(\theta,\psi)=\frac{m^2}{2({\mathsf I}\tilde{\bm C}_3)\cdot\tilde{\bm C}_3}+V(\theta,\psi).
\end{equation}

Relative equilibria with $\dot\theta=0$, $\dot\psi=0$ are found from the effective
potential \eqref{eqnnlr68} solving the equations
\begin{subequations}\label{eqnnlr69}
\begin{align}
\frac{\partial V_m}{\partial\theta}&=-\frac{m^2({\mathsf I}\tilde{\bm C}_3)}{[({\mathsf I}\tilde{\bm C}_3)\cdot\tilde{\bm C}_3]^2}\cdot\frac{\partial\tilde{\bm C}_3}
{\partial\theta}+\frac{\partial V}{\partial\theta}=0,\label{eqnnlr69.1}\\
\frac{\partial V_m}{\partial\psi}&=-\frac{m^2({\mathsf I}\tilde{\bm C}_3)}{[({\mathsf I}\tilde{\bm C}_3)\cdot\tilde{\bm C}_3]^2}\cdot\frac{\partial\tilde{\bm C}_3}
{\partial\psi}+\frac{\partial V}{\partial\psi}=0,\label{eqnnlr69.2}
\end{align}
\end{subequations}
with
\begin{subequations}\label{eqnnlr70}
\begin{align}
\frac{\partial\tilde{\bm C}_3}{\partial\theta}&=(-\cos\psi\cos\theta,\sin\psi\cos\theta,-\sin\theta),\label{eqnnlr70.1}\\
\frac{\partial\tilde{\bm C}_3}{\partial\psi}&=(\sin\psi\sin\theta,\cos\psi\sin\theta,0) \label{eqnnlr70.2}
\end{align}
\end{subequations}
and
\begin{subequations}\label{eqnnlr71}
\begin{align}
\frac{\partial V}{\partial\theta}
&=\omega\sin\theta-2\Delta\omega_2\sin\theta\cos\theta\sin^2\psi+2\Delta\omega_3\sin\theta\cos\theta,
\label{eqnnlr71.1}\\
\frac{\partial V}{\partial\psi}&=-2\Delta\omega_2\sin\psi\cos\psi\sin^2\theta.\label{eqnnlr71.2}
\end{align}
\end{subequations}

In general the solutions of equation \eqref{eqnnlr69} depend on $m$.
For a given $m$ the zero set of $\frac{\partial V_m}{\partial\theta}$
(respectively, $\frac{\partial V_m}{\partial\psi}$) in the $\theta$-$\psi$ space
gives the solution set for equation \eqref{eqnnlr69.1} (respectively, \eqref{eqnnlr69.2}).
The solution set typically consists of (possibly disjoint)
curves in the $\theta$-$\psi$ plane.
In practice (see below), we are able to give a natural parametrization of each of these solution
curves, so that
the general solution of equations \eqref{eqnnlr69} is obtained by evaluating
$\frac{\partial V_m}{\partial\psi}$ (respectively $\frac{\partial V_m}{\partial\theta}$) along each of the
particular solution curves. Zeroes of the relevant functions are found by interpolation along the
curve, yielding values of $\theta$ and $\psi$ that solve \eqref{eqnnlr69} for given $m$.
Finally, the effective potential \eqref{eqnnlr68} is evaluated
at each of these solution points, which  in general have different energies for a given $m$.
The calculation is repeated for different values of $m$ to obtain the complete $E$-$m$ diagram.

To clarify the general procedure just outlined, consider
the Euler problem (free asymmetric top), with
$\omega=\Delta\omega_2=\Delta\omega_3=0$.  Equations \eqref{eqnnlr69} in this case are
\begin{subequations}\label{eqnnlr72}
\begin{align}
\frac{\partial V_m}{\partial\theta}&=-\frac{\sin\theta\cos\theta
\left(I_1\cos^2\psi+I_2\sin^2\psi-I_3\right)m^2}{[({\mathsf I}\tilde{\bm C}_3)\cdot\tilde{\bm C}_3]^2}=0,
\label{eqnnlr72.1}\\
\frac{\partial V_m}{\partial\psi}&=-\frac{\sin\psi\cos\psi\sin^2\theta \,(I_2-I_1)m^2}
{[({\mathsf I}\tilde{\bm C}_3)\cdot\tilde{\bm C}_3]^2}=0.\label{eqnnlr72.2}
\end{align}
\end{subequations}
The common denominator is never zero, and is sufficient to find the zeros in the numerators.
The solutions $\theta=0,\pi$ satisfy both equations simultaneously
giving the solution set $\mathcal{A}=\{0,\pi\}\times[0,2\pi)$.
In the cartesian product the first set gives the possible values of the $\theta$ coordinate,
the second set the possible values of the $\psi$ coordinate.

Setting $\psi=0$, $\pi/2$, $\pi$, or $3\pi/2$ solves Eq.\ \eqref{eqnnlr72.2};
substituting these values
into \eqref{eqnnlr72.1} gives\begin{subequations}\label{eqnnlr73}
\begin{align}
\sin\theta\cos\theta\left(I_3-I_1\right)m^2&=0,\quad \mbox{$\psi=0,\pi$},\label{eqnnlr73.1}\\
\sin\theta\cos\theta\left(I_3-I_2\right)m^2&=0,\quad \mbox{$\psi=\pi/2,3\pi/2$}.\label{eqnnlr73.2}
\end{align}
\end{subequations}
For these equations $\theta=0,\pi/2,\pi$, are solutions.
In the same notation used before the new solutions are $\mathcal{B}=\{\pi/2\}\times\{0,\pi\}$,
and $\mathcal{C}=\{\pi/2\}\times\{\pi/2,3\pi/2\}$.

The complete solution set for the Euler problem is
$\mathcal{S}_E=\mathcal{A}\cup\mathcal{B}\cup\mathcal{C}$.
For each of these sets is possible to obtain a $V_m$-$m$ curve simply by evaluating
the effective potential on each of the solution sets
\begin{subequations}\label{eqnnlr76}
\begin{align}
V_m(\mathcal{A})=\frac{m^2}{2I_3},\label{eqnnlr76.1}\\
V_m(\mathcal{B})=\frac{m^2}{2I_1},\label{eqnnlr76.2}\\
V_m(\mathcal{C})=\frac{m^2}{2I_2}.\label{eqnnlr76.3}
\end{align}
\end{subequations}

The $E$-$m$ diagram consists of three parabolas and the regions enclosed between them.
This diagram is shown in Fig.\ \ref{fignlr19} for the three molecules considered here
(cf.\ \S 3.4 of Ref.\ \onlinecite{Arnold88}; Ch.\ 14 of Ref.\ \onlinecite{Bolsinov04}).
The different lines represent the three parabolas \eqref{eqnnlr76} and correspond physically
to rotations of the top about the body fixed axes
(in a positive or negative sense). The region below the red curve is physically inaccessible.
In the axis convention used to obtain the asymmetric top Hamiltonian the red curve,
equation \eqref{eqnnlr76.2}, corresponds to stable rotation about the body fixed $x$-axis;
the green parabola, equation \eqref{eqnnlr76.3}, to an unstable rotation about the $y$-axis,
and the blue curve to a stable rotation about the $z$-axis.

For a given ($E$,$m$) point in Fig.\ \ref{fignlr19}, the effective potential \eqref{eqnnlr68}
gives an equation to solve in $\theta$ or $\psi$. The solutions are found as contours
of constant $V_m=E$ in the $\theta$-$\psi$ configuration space, the Poisson sphere $S^2$.
These contours define the classically accessible ($V_m \leqslant E$) and forbidden ($V_m>E$) regions
for given $E$ and $m$. The topology of the different solutions on the
$(\theta, \psi)$ sphere is characterized by
the (multi-valued) \emph{genus} \cite{Arnold88}, which for every connected classically allowed region
counts the number of disjoint discs associated with forbidden motion that are removed from the sphere.

Figure \ref{fignlr20} shows the classically accessible regions (in black) of $S^2$ for
iodopentafluorobenzene. Panel \ref{fignlr20}(a) represents an ($E$,$m$) point located between the
red and green parabolas in Fig.\ \ref{fignlr19}(c);
the multivalued genus is \emph{1,1} since there are two disjoint classically accessible
regions and from the point of view of each there is one white disc removed from the sphere.
In the same way, panel \ref{fignlr20}(b) is the accessible region for a ($E$,$m$) point
located between the green and blue parabolas of \ref{fignlr19}(c), but now the genus is \emph{2} since
two white discs are removed. Finally the region above the blue parabola in \ref{fignlr19}(c)
has \emph{0} genus since all the sphere is classically accessible (panel \ref{fignlr20}(c)).

\newpage

\section{$E$-$m$ diagram for asymmetric top molecules in static electric fields}
\label{em_electric}

Energy-momentum diagrams for the asymmetric top in an external gravitational field have been
studied by Katok \cite{Katok72} and by Gashenenko and Richter \cite{Gashenenko04}.
An important conclusion in these works is that, in the study of relative equilibria and their
bifurcations in asymmetric tops, the relevant parameters are the ratios between two of the
moments of inertia and the third one, \emph{e.g.}, $I_1/I_3$ and $I_2/I_3$.

In this section we treat as examples three asymmetric top molecules of recent theoretical
and experimental interest \cite{Kozin03,Stapelfeldt03}: the near-prolate top iodobenzene
(C$_6$H$_5$I) \cite{Bulthuis97,Poulsen04},
the near-oblate top pyridazine (C$_4$H$_4$N$_2$) \cite{Kozin03,Li98},
and the intermediate case iodopentafluorobenzene \cite{Poulsen04}.
Relevant physical  parameters for these molecules are given in Table \ref{tabnlr1}.

In Fig.\ \ref{fignlr20a} we show the definition of the body fixed frame for these molecules.
The moment of inertia $I_i$ is related to the rotational constant $B_i$ by $B_i=(2I_i)^{-1}$.
The rotational constants $B_i$, the field parameter $\omega$,
and the energy are all scaled by $B_3$. For iodobenzene an electric field of $\varepsilon_S=25 \,
\mathrm{kVcm^{-1}}$ \cite{Bulthuis97} gives $\omega/B_3=4.52$; for pyridazine a field of strength
$\varepsilon_S=56 \, \mathrm{kVcm^{-1}}$ \cite{Li98} gives $\omega/B_3=19.04$; in iodopentafluorobenzene, an electric field of $\varepsilon_S=25 \,\mathrm{kVcm^{-1}}$ gives  $\omega/B_3=18.93$ \cite{Bulthuis97,Kozin03,Poulsen04}. 

For the asymmetric top molecule in an electric field, the relative equilibria are found by
solving the equations (setting $\Delta \omega_2 = \Delta\omega_3 =0$ in Equations \eqref{eqnnlr69}):
\begin{subequations}\label{eqnnlr77}
\begin{align}
\frac{\partial V_m}{\partial\theta}&=
-\frac{\sin\theta\cos\theta\left(I_1\cos^2\psi+I_2\sin^2\psi-I_3\right)m^2}
{[({\mathsf I}\tilde{\bm C}_3)\cdot\tilde{\bm C}_3]^2}+\omega\sin\theta=0,\label{eqnnlr77.1}\\
\frac{\partial V_m}{\partial\psi}&=-\frac{\sin\psi\cos\psi\sin^2\theta(I_2-I_1)m^2}{[({\mathsf I}\tilde{\bm C}_3)\cdot\tilde{\bm C}_3]^2}=0,\label{eqnnlr77.2}
\end{align}
\end{subequations}
where
\begin{equation}\label{eqnnlr78}
({\mathsf I}\tilde{\bm C}_3)\cdot\tilde{\bm C}_3=I_1\sin^2\theta\cos^2\psi+I_2\sin^2\theta\sin^2\psi+I_3\cos^2\theta.
\end{equation}
Again we have the  solution set $\mathcal{A}=\{0,\pi\}\times\left[0,2\pi\right)$.
There are two solution subsets $\mathcal{A}_1\subset\mathcal{A}$, and $\mathcal{A}_2\subset\mathcal{A}$,
given by $\mathcal{A}_1=\{0\}\times\left[0,2\pi\right)$,
and $\mathcal{A}_2=\{\pi\}\times\left[0,2\pi\right)$. The $V_m$-$m$ curves for these subsets are
\begin{subequations}\label{eqnnlr79}
\begin{align}
V_m(\mathcal{A}_1)&=\frac{m^2}{2I_3}-\omega,\label{eqnnlr79.1}\\
V_m(\mathcal{A}_2)&=\frac{m^2}{2I_3}+\omega.\label{eqnnlr79.2}
\end{align}
\end{subequations}

The solutions $\psi=0, \pi/2, \pi, 3\pi/2$ to \eqref{eqnnlr77.2} must be
substituted into \eqref{eqnnlr77.1}. After rearranging and dividing by $I_3$, this gives
\begin{subequations}\label{eqnnlr80}
\begin{align}
\left(1-i_1\right)m^2\cos\theta&=-\omega I_3\left[i_1+\left(1-i_1\right)\cos^2\theta\right]^2,\label{eqnnlr80.1}\\
\left(1-i_2\right)m^2\cos\theta&=-\omega I_3\left[i_2+\left(1-i_2\right)\cos^2\theta\right]^2,\label{eqnnlr80.2}
\end{align}
\end{subequations}
for $\psi=0,\pi$ and $\psi=\pi/2,3\pi/2$ respectively, and $i_1\equiv I_1/I_3$, $i_2\equiv I_2/I_3$.
For the case $I_1>I_2>I_3$, \emph{i.e.}, $i_1>i_2>1$, the right hand sides (RHS) of
these equations are functions with maxima at $\theta=0,\pi$ and a minimum at $\theta=\pi/2$;
the left hand sides (LHS) are functions with a minimum at $\theta=0$, a maximum at $\theta=\pi$,
and a fixed zero at $\theta=\pi/2$.
In Fig.\ \ref{fignlr21}, the LHS and RHS of these equations are plotted in
order to show the nature of the solutions.
As $m$ is increased the amplitude of variation of the solid curve about zero gets larger until
it intersects the dashed line at $\theta=0$, at which point a solution of the equation is obtained.
Since the curves do not change their shape, there can exist only one solution for each of
the equations \eqref{eqnnlr80} for a given $m$. This unique intersection gives a value of $\theta$,
which together with the respective $\psi$ gives the solution of equations \eqref{eqnnlr80}.
The form of the curves indicates that the intersection occurs initially at $\theta=0$ and
then moves to larger values of $\theta$ as  $|m|$ increases, \emph{i.e.}, the solution
$\mathcal{A}_1$ bifurcates twice, first for the $\psi=0,\pi$ curve and
then for the $\psi=\pi/2,3\pi/2$ curve.
Since the solid line is always zero at $\theta=\pi/2$,
the intersection of the two curves cannot go beyond this point, this means that for large
values of $m$ the asymmetric top in an electric field behaves like an Euler top.

The bifurcations in the $E$-$m$ and $E$-$\theta$ diagrams are shown in Fig.\ \ref{fignlr22}.
For iodobenzene we set  $\omega/B_3=10$, for pyridazine $\omega/B_3=20$, and for
iodopentafluorobenzene $\omega/B_3=20$.

In the $E$-$m$ diagrams there are four regions delimited by different color curves.
The allowed configuration space corresponding to the regions delimited by the red
and green curves have genus \emph{1,1};
those between the green and the blue curves have genus \emph{2};
between the blue and magenta curves the genus is \emph{1}, and above the magenta curve the genus \emph{0}.
It can be seen that the only difference between these diagrams and those for
the Euler top is the presence of the region with genus \emph{1}.

The curves themselves represent rotations in $\phi$ at constant $\theta$ and $\psi$.
Associated $\theta$ values for given energy are shown using the curve colors in
panels \ref{fignlr22}(b), (d), and (f).
The red curve is a stable rotation with $\psi=0,\pi$ and $\theta$; the green curve is
unstable with $\psi=\pi/2,3\pi/2$ and $\theta$ in the RHS panels.
The blue and magenta curves are associated with degenerate equilibria in the $\psi$ coordinate and $\theta=0$
(stable, blue) or $\theta=\pi$ (unstable, magenta). Physically the blue and magenta curves in Fig.\
\ref{fignlr22} represent the situation when the molecule's dipole is oriented with
the field (blue) and against it (magenta), while at the same time the molecule is rotating about its
own $z$-axis with direction and angular velocity given by $m$.

\newpage

\section{$E$-$m$ diagrams for asymmetric top molecules in nonresonant laser fields}
\label{em_laser}

The effective potential for interaction of a molecule with a nonresonant linearly polarized
laser field ($\omega = 0$) is
\begin{equation}\label{eqnnlr81}
V_m(\theta,\psi)=\frac{m^2}{2({\mathsf I}\tilde{\bm C}_3)\cdot\tilde{\bm C}_3}-\Delta\omega_2\sin^2\theta \sin^2\psi - \Delta\omega_3\cos^2\theta.
\end{equation}
The relative equilibria are obtained by solving
\begin{subequations}\label{eqnnlr82}
\begin{align}
\frac{\partial V_m}{\partial\theta}&=-\frac{\sin\theta\cos\theta\left(I_1\cos^2\psi+I_2\sin^2\psi-I_3\right)m^2}{[({\mathsf I}\tilde{\bm C}_3)\cdot\tilde{\bm C}_3]^2}
+\frac{\partial V}{\partial\theta}=0,\label{eqnnlr82.1}\\
\frac{\partial V_m}{\partial\psi}&=-\frac{\sin\psi\cos\psi\sin^2\theta(I_2-I_1)m^2}{[({\mathsf I}\tilde{\bm C}_3)\cdot\tilde{\bm C}_3]^2}+\frac{\partial V}{\partial\psi}=
0,\label{eqnnlr82.2}
\end{align}
\end{subequations}
where $({\mathsf I}\tilde{\bm C}_3)\cdot\tilde{\bm C}_3$ is again given by Eq.\ \eqref{eqnnlr78}
and
\begin{subequations}\label{eqnnlr84}
\begin{align}
\frac{\partial V}{\partial\theta}&=-2\Delta\omega_2\sin\theta\cos\theta\sin^2\psi+2\Delta\omega_3\sin\theta\cos\theta,\label{eqnnlr84.1}\\
\frac{\partial V}{\partial\psi}&=-2\Delta\omega_2\sin\psi\cos\psi\sin^2\theta.\label{eqnnlr84.2}
\end{align}
\end{subequations}

From \eqref{eqnnlr82}-\eqref{eqnnlr84} it is seen that $\theta=0,\pi$ are again simultaneous
solutions and the first solution set is $\mathcal{A}=\{0,\pi\}\times\left[0,2\pi\right)$.
In contrast with the electric field case, the symmetry of the laser interaction gives
only one $V_m$-$m$ curve for $\mathcal{A}$,
\begin{equation}\label{eqnnlr85}
V_m(\mathcal{A})=\frac{m^2}{2I_3}-\Delta\omega_3.
\end{equation}
It is straightforward to see that $\theta=\pi/2$ is a solution of  \eqref{eqnnlr82.1},
which after substitution into \eqref{eqnnlr82.2} gives
\begin{equation}\label{eqnnlr86}
-\frac{\sin\psi\cos\psi(I_2-I_1)m^2}{I_1\cos^2\psi+I_2\sin^2\psi}-2\Delta\omega_2\sin\psi\cos\psi=0,
\end{equation}
with immediate solutions $\psi=0,\pi/2,\pi,3\pi/2$.
The second solution set is therefore $\mathcal{B}=\{\pi/2\}\times\{0,\pi/2,\pi,3\pi/2\}$.
There are two subsets, $\mathcal{B}_1\subset\mathcal{B}$
and $\mathcal{B}_2\subset\mathcal{B}$, given by $\mathcal{B}_1=\{\pi/2\}\times\{0,\pi\}$
and $\mathcal{B}_2=\{\pi/2\}\times\{\pi/2,3\pi/2\}$. For these the $V_m$-$m$ curves are
\begin{subequations}\label{eqnnlr87}
\begin{align}
V_m(\mathcal{B}_1)&=\frac{m^2}{2I_1},\label{eqnnlr87.1}\\
V_m(\mathcal{B}_2)&=\frac{m^2}{2I_2}-\Delta\omega_2.\label{eqnnlr87.2}
\end{align}
\end{subequations}

Now,  $\psi=0$, $\pi/2$, $\pi$, and $3\pi/2$ are particular solutions of
equation \eqref{eqnnlr82.2}; substituting these
values into equation \eqref{eqnnlr82} and dividing by $I_3$ produces
\begin{subequations}\label{eqnnlr88}
\begin{align}
\left(1-i_1\right)m^2&=-2\Delta\omega_3 I_3\left[i_1+\left(1-i_1\right)\cos^2\theta\right]^2,\label{eqnnlr88.1}\\
\left(1-i_2\right)m^2&=2(\Delta\omega_2-\Delta\omega_3)I_3\left[i_2+\left(1-i_2\right)\cos^2\theta\right]^2,\label{eqnnlr88.2}
\end{align}
\end{subequations}
for $\psi=0$, $\pi$, and $\psi=\pi/2$, $3\pi/2$ respectively, and as before
$i_1\equiv I_1/I_3$ and $i_2\equiv I_2/I_3$, with $i_1>i_2>1$.
The nature of the solutions depends on the values of $\Delta\omega_2$ and $\Delta\omega_3$.
For iodobenzene and pyridazine $\Delta\omega_3>\Delta\omega_2>0$, since for
these molecules $\alpha_1<\alpha_2<\alpha_3$ (cf.\ Table \ref{tabnlr1}).
This is however not always the case: for pyridine
with the same axis convention $\alpha_1<\alpha_3<\alpha_2$ \cite{Hinchliffe94},
which implies $\Delta\omega_2>\Delta\omega_3>0$.

For a physically reasonable laser intensity $10^{12}\,\mathrm{Wcm^{-2}}$ applied to
iodobenzene we have $\Delta\omega_2/B_3=284.6$ and $\Delta\omega_3/B_3=630.5$;
the same field applied to pyridazine produces $\Delta\omega_2/B_3=225.7$ and $\Delta\omega_3/B_3=228.9$; iodobenzene gives $\Delta\omega_2/B_3=2280.06$ and $\Delta\omega_3/B_3=4097.94$. These dimensionless ratios are much larger than the corresponding energy
ratios for physically relevant values of the interaction with a static electric field,
$\omega=10$ and $\omega=20$.
As we wish to investigate the interesting dynamical regime in which the
effects of both fields are of similar magnitude (see next Section),
we reduce the intensity of the laser by a factor of $10$ to obtain the values listed in Table \ref{tabnlr2}.

The plot of the RHS and LHS of
equations \eqref{eqnnlr88} is shown in Fig.\ \ref{fignlr23}.
The situation is different from the static electric field case.
Now the LHS of the equations, the solid horizontal line, moves down as $m$ increases.
This line intersects the RHS curve (dashed) at $\theta=0,\pi$ simultaneously for $m$ values
\begin{subequations}\label{eqnnlr89}
\begin{align}
M_1&=\pm\left[\frac{-2\Delta\omega_3 I_3}{1-i_1}\right]^{1/2},\label{eqnnlr89.1}\\
M_2&=\pm\left[\frac{2(\Delta\omega_2-\Delta\omega_3) I_3}{1-i_2}\right]^{1/2},\label{eqnnlr89.2}
\end{align}
\end{subequations}
for $\psi=0,\pi$ and $\psi=\pi/2,3\pi/2$, respectively.
As the value of $m$ increases the points of
intersection approach the value $\theta=\pi/2$ symmetrically, \emph{i.e.},
$\pi/2-\theta_{\mathrm{left}}=\theta_{\mathrm{right}}-\pi/2$.
Finally, at $m$ values
\begin{subequations}\label{eqnnlr90}
\begin{align}
\overline{M}_1&=\pm M_1 i_1,\label{eqnnlr90.1}\\
\overline{M}_2&=\pm M_2 i_2,\label{eqnnlr90.2}
\end{align}
\end{subequations}
for $\psi=0,\pi$ and $\psi=\pi/2,3\pi/2$, respectively, the intersection occurs exactly at $\theta=\pi/2$.
For larger values of $m$ there is no solution.

Considering only positive values of $m$, equations \eqref{eqnnlr88} can be solved for $\cos^2\theta$
to get
\begin{subequations}\label{eqnnlr91}
\begin{align}
\cos^2\theta=\frac{m-\overline{M}_1}{M_1-\overline{M}_1},\label{eqnnlr91.1}\\
\cos^2\theta=\frac{m-\overline{M}_2}{M_2-\overline{M}_2},\label{eqnnlr91.2}
\end{align}
\end{subequations}
for $\psi=0,\pi$ and $\psi=\pi/2,3\pi/2$, respectively.
The two solutions of each of these equations together with the respective $\psi$ values
give the solution sets $\mathcal{C}$ for  $\psi=0,\pi$ and $\mathcal{D}$ for $\psi=\pi/2,3\pi/2$.
The $V_m$-$m$ curves for these solutions are
\begin{subequations}\label{eqnnlr92}
\begin{align}
V_m(\mathcal{C})=&\frac{(M_1-\overline{M}_1)m^2}
{2\left[I_3(m-\overline{M}_1)+I_1(M_1-m)\right]}
-\Delta\omega_3\frac{m-\overline{M}_1}{M_1-\overline{M}_1},\label{eqnnlr92.1}\\
V_m(\mathcal{D})=&\frac{(M_2-\overline{M}_2)m^2}
{2\left[I_3(m-\overline{M}_2)+I_2(M_2-m)\right]}
-\Delta\omega_2+(\Delta\omega_2-\Delta\omega_3)\frac{m-\overline{M}_2}{M_2-\overline{M}_2}\label{eqnnlr92.2},
\end{align}
\end{subequations}
with $m\in\left[M_1,\overline{M}_1\right]$ and $m\in\left[M_2,\overline{M}_2\right]$ for the first and
second equation respectively.
The value of $V_m(\mathcal{C})$ at $m=M_1$ is equal to that of $V_m(\mathcal{A})$ at $m=M_1$,
while its value at $m=\overline{M}_1$ is equal to $V_m(\mathcal{B}_1)$ at $m=\overline{M}_1$;
similarly, the value of $V_m(\mathcal{D})$ at $m=M_2$ equals $V_m(\mathcal{A})$ at $m=M_2$,
and at $m=\overline{M}_2$ equals $V_m(\mathcal{B}_2)$ at $m=M_2$.
These results indicate that in the $E$-$m$ diagram the solution set $\mathcal{C}$ is a bridge
connecting solutions $\mathcal{A}$ and $\mathcal{B}_1$.  In like fashion
the solution set $\mathcal{D}$ connects the solutions $\mathcal{A}$ and $\mathcal{B}_2$.

Finally, the last solution set is obtained from equations \eqref{eqnnlr82} after removing all the common factors
\begin{subequations}\label{eqnnlr93}
\begin{align}
-\frac{\left(I_1\cos^2\psi+I_2\sin^2\psi-I_3\right)m^2}{\left(I_1\sin^2\theta\cos^2\psi+I_2\sin^2\theta\sin^2\psi+I_3\cos^2\theta\right)
^2}-2\Delta\omega_2\sin^2\psi+2\Delta\omega_3&=0,\label{eqnnlr93.1}\\
-\frac{(I_2-I_1)m^2}{\left(I_1\sin^2\theta\cos^2\psi+I_2\sin^2\theta\sin^2\psi+I_3\cos^2\theta\right)^2}-2\Delta\omega_2&=0.\label{eqnnlr93.2}
\end{align}
\end{subequations}
Solutions of these equations for given $m$
are obtained by finding the intersections of the zero contours of the LHS of
these equations in the $\theta$-$\psi$ space.
For the case of the laser field, these equations can be rearranged to get $m^2$ in terms of $\theta$ and $\psi$
\begin{subequations}\label{eqnnlr94}
\begin{align}
m^2&=\frac{2(\Delta\omega_2\sin\psi^2-\Delta\omega_3)}{I_3-I_1\cos^2\psi-I_2\sin^2\psi}\left[({\mathsf I}\tilde{\bm C}_3)\cdot\tilde{\bm C}_3\right]
^2,\label{eqnnlr94.1}\\
m^2&=\frac{2\Delta\omega_2}{I_1-I_2}\left[({\mathsf I}\tilde{\bm C}_3)\cdot\tilde{\bm C}_3\right]^2,\label{eqnnlr94.2}
\end{align}
\end{subequations}
where the definition \eqref{eqnnlr78} is used.
With the equations written in this form is clear that we must find $\psi$ such that
\begin{equation}\label{eqnnlr95}
\frac{2(\Delta\omega_2\sin^2\psi-\Delta\omega_3)}{I_3-I_1\cos^2\psi-I_2\sin^2\psi}=\frac{2\Delta\omega_2}{I_1-I_2}.
\end{equation}
Rearranging and simplifying gives the condition
\begin{equation}\label{eqnnlr96}
1-\frac{\Delta\omega_3}{\Delta\omega_2}=\frac{I_3-I_2}{I_1-I_2},
\end{equation}
which can be written in terms of the polarizability
\begin{equation}\label{eqnnlr97}
\frac{\alpha_3-\alpha_2}{\alpha_1-\alpha_2}=\frac{I_3-I_2}{I_1-I_2}.
\end{equation}
In general this equality is not satisfied for physical parameter values.
For example, table \ref{tabnlr2} lists the LHS and RHS of equation \eqref{eqnnlr97} for the molecules of interest.


As condition \eqref{eqnnlr96} is not fulfilled for the molecules considered here,
the last solution set is empty,
$\mathcal{E}=\emptyset$.
The complete solution for the asymmetric top rotor in a linearly polarized laser field is then
given by $\mathcal{S}_L=\mathcal{A}\cup\mathcal{B}\cup\mathcal{C}\cup\mathcal{D}$.
The $V_m$-$m$ curves for these solutions in the case of iodopentafluorobenzene, iodobenzene
and pyridazine are shown in
Figs \ref{fignlr24.0}, \ref{fignlr24}, and \ref{fignlr25}, respectively.
In these figures the solution sets are given by different types of curves.
The solution set $\mathcal{A}$ is given by the red curve, $\mathcal{B}_1$ by the green curve
and $\mathcal{B}_2$ by the blue. The bridge solution sets $\mathcal{C}$,
connecting $\mathcal{A}$ with $\mathcal{B}_1$ are magenta; the bridge $\mathcal{D}$,
connecting $\mathcal{A}$ with $\mathcal{B}_2$, is the cyan curve.

For iodopentafluorobenzene and iodobenzene there are nine different regions
delimited by different curve types with their corresponding genera indicated with arrows in Fig.\ \ref{fignlr24}.
The classically accessible $\theta$-$\psi$ configuration space for each of these regions is
shown in Fig.\ \ref{fignlr26}. The simplest region is characterized by genus \emph{0};
in this region the molecule is free to move in any possible configuration in $\theta$-$\psi$
as can be seen in Fig.\ \ref{fignlr26}(i).
There are two regions with genus \emph{2}: one for high energy and large $m$, the other for
low energy and small $m$. In the high energy region \ref{fignlr26}(a) shows
that the molecule is localized in the equatorial
region and the poles are forbidden. The second region with genus \emph{2} is shown in \ref{fignlr26}(c);
in this case the forbidden regions are on the equator with $\psi=0,\pi$.
For genus \emph{4} there is only one region, shown in panel \ref{fignlr26}(f);
the molecule can access neither the poles nor the
equatorial regions with $\psi=0,\pi$. In the case of genus \emph{2,2}, panel \ref{fignlr26}(d),
the accessible region is delocalized in $\psi$ and restricted in $\theta$ to a region close,
but not on, the poles. Finally, for genus \emph{1,1,1,1} (\ref{fignlr26}(f))
the motion is highly localized near $\theta=\pi/4, 3\pi/4$ and $\psi=\pi/2,3\pi/2$.

The fact that both the inertia and polarizability tensors of pyridazine have near-oblate symmetry
makes it harder to observe the different regions in the $E$-$m$ diagram,
as can be seen from Fig.\ \ref{fignlr25}. As $I_2 \approx I_3$
and $\Delta\omega_2 \approx \Delta\omega_3$, the $V_m$-$m$ curves for the
solution sets $\mathcal{A}$ and $\mathcal{B}_2$ are very similar.
The magenta bridging solution $\mathcal{C}$ connecting $\mathcal{A}$ and $\mathcal{B}_1$ is
easily seen, but the cyan $\mathcal{D}$ bridge is much harder to see.
As for iodopentafluorobenzene and iodobenzene, there are 9 different regions which have the same distribution of genus.

\newpage

\section{$E$-$m$ diagrams for asymmetric top molecules in collinear fields}
\label{em_collinear}

We now consider the rotational dynamics of dipolar asymmetric tops in combined static
(Section \ref{em_electric}) and nonresonant laser (Section \ref{em_laser}) fields.
The polarization of the laser field is taken to be collinear with the static
field, so that $m$ is a conserved quantity.

A related problem has been analyzed by Tatarinov \cite{Tatarinov74,Arnold88},
who studied the problem of the rotation of a rigid body about a fixed point with a potential
\begin{equation}\label{eqnnlr98}
V=P\tilde{\bm C}_3\cdot{\bm R}_{CM}+\frac{\rho}{2}({\mathsf I}\tilde{\bm C}_3)\cdot\tilde{\bm C}_3,
\end{equation}
where $P$ and $\rho$ are constants, and ${\bm R}_{CM}$ is the location of
the center of mass relative to the fixed point.
In Tatarinov's work $E$-$m$ diagrams and genera are obtained for the effective potential.

In molecular terms, the first term of the potential \eqref{eqnnlr98}
corresponds to the interaction of an electric field along the space-fixed $z$-axis
with the dipole moment of the molecule, where the dipole moment vector can point in an
arbitrary direction in the molecule-fixed frame, $\boldsymbol{R}$, not only along the molecule fixed $z$-axis
as in the case studied here.
The second term of \eqref{eqnnlr98} has no obvious molecular analog,
but the dependence on Euler angles is exactly the same as for the laser interaction.
In fact, the potential energy for an asymmetric top in tilted (non-collinear)
fields can be written
\begin{equation}\label{eqnnlr99}
V=-\bm{\Omega}\cdot\tilde{\bm C}_3-({\Delta\Omega}\tilde{\bm C}_3)\cdot\tilde{\bm C}_3,
\end{equation}
with the definitions $\bm{\Omega}\equiv\left(\omega\sin\beta,0,\omega\cos\beta\right)$,
and  ${\Delta\Omega}\equiv\mathrm{diag}\left(0,\Delta\omega_2,\Delta\omega_3\right)$.

The effective potential for the dipolar asymmetric top in collinear fields, where the dipole moment
points along the molecule-fixed $z$-axis, is (cf.\ eq.\ \eqref{eqnnlr68})
\begin{equation}\label{eqnnlr100}
V_m(\theta,\psi)=\frac{m^2}{2({\mathsf I}\tilde{\bm C}_3)\cdot\tilde{\bm C}_3}-\omega\cos\theta-\Delta\omega_2\sin^2\theta \sin^2\psi - \Delta\omega_3\cos^2\theta.
\end{equation}
The relative equilibria are obtained by solving the equations
\begin{subequations}\label{eqnnlr101}
\begin{align}
\frac{\partial V_m}{\partial\theta}&=
-\frac{\sin\theta\cos\theta\left(I_1\cos^2\psi+I_2\sin^2\psi-I_3\right)m^2}
{[({\mathsf I}\tilde{\bm C}_3)\cdot\tilde{\bm C}_3]^2}+\frac{\partial V}{\partial\theta}=0,\label{eqnnlr101.1}\\
\frac{\partial V_m}{\partial\psi}&=-\frac{\sin\psi\cos\psi\sin^2\theta(I_2-I_1)m^2}{[({\mathsf I}\tilde{\bm C}_3)\cdot\tilde{\bm C}_3]^2}+\frac{\partial V}{\partial\psi}=
0,\label{eqnnlr101.2}
\end{align}
\end{subequations}
with
\begin{equation}\label{eqnnlr102}
({\mathsf I}\tilde{\bm C}_3)\cdot\tilde{\bm C}_3=I_1\sin^2\theta\cos^2\psi+I_2\sin^2\theta\sin^2\psi+I_3\cos^2\theta,
\end{equation}
and
\begin{subequations}\label{eqnnlr103}
\begin{align}
\frac{\partial V}{\partial\theta}&=\omega\sin\theta-2\Delta\omega_2\sin\theta\cos\theta\sin^2\psi+2\Delta\omega_3\sin\theta\cos\theta,\label{eqnnlr103.1}\\
\frac{\partial V}{\partial\psi}&=-2\Delta\omega_2\sin\psi\cos\psi\sin^2\theta.\label{eqnnlr103.2}
\end{align}
\end{subequations}

In both equations \eqref{eqnnlr101} the common factor $\sin\theta$ gives the simultaneous
solution $\theta=0$ or $\pi$, so the first solution set
is $\mathcal{A}=\{0,\pi\}\times\left[0,2\pi\right)$.
As for the static electric field case, this solution generates two subsets
$\mathcal{A}_1=\{0\}\times\left[0,2\pi\right)$
and $\mathcal{A}_2=\{\pi\}\times\left[0,2\pi\right)$. The $V_m$-$m$ curves for these solutions are
\begin{subequations}\label{eqnnlr104}
\begin{align}
V_m(\mathcal{A}_1)&=\frac{m^2}{2I_3}-\omega-\Delta\omega_3,\label{eqnnlr104.1}\\
V_m(\mathcal{A}_2)&=\frac{m^2}{2I_3}+\omega-\Delta\omega_3.\label{eqnnlr104.2}
\end{align}
\end{subequations}
As for the laser interaction case, equation \eqref{eqnnlr101.2} has solutions
$\psi=0$, $\pi/2$, $\pi$, and $3\pi/2$, which after substitution into \eqref{eqnnlr101.1} give
\begin{subequations}\label{eqnnlr105}
\begin{align}
\cos\theta\left(1-i_1\right)m^2&=-I_3(\omega+2\Delta\omega_3\cos\theta)\left[i_1+\left(1-i_1\right)\cos^2\theta\right]^2,\label{eqnnlr105.1}\\
\cos\theta\left(1-i_2\right)m^2&=-I_3\left[\omega-2(\Delta\omega_2-\Delta\omega_3)\cos\theta\right]\left[i_2+\left(1-i_2\right)\cos^2\theta\right]^2.\label{eqnnlr105.2}
\end{align}
\end{subequations}

Figures \ref{fignlr28.0}, \ref{fignlr28} and \ref{fignlr29} show plots of
the  LHS and RHS of equations \eqref{eqnnlr105} for iodopentafluorobenzene ($\omega/B_3=20$,
$\Delta\omega_2/B_3=228.006$ and $\Delta\omega_3/B_3=409.794$), iodobenzene ($\omega/B_3=10$,
$\Delta\omega_2/B_3=28.456$ and $\Delta\omega_3/B_3=63.05$) and for pyridazine
($\omega/B_3=20$, $\Delta\omega_2/B_3=22.56$, and $\Delta\omega_3/B_3=22.88$).

Figures \ref{fignlr28.0} and \ref{fignlr28} show the LHS of \eqref{eqnnlr105} for iodopentafluorobenzene
and iodobenzene (respectively) with $m=0$ as a dotted line in both cases;
it is clear that on each figure that for $m=0$ there is only one intersection at $\theta>\pi/2$.
As the value of $m$ increases, the dashed curve will intersect the solid line at additional points.
The RHS of \eqref{eqnnlr105.1} gives $-I_3\left(\omega+2\Delta\omega_3\right)$ at $\theta=0$
and $-I_3\left(\omega-2\Delta\omega_3\right)$ at $\theta=\pi$; for these two melecules in the
fields specified above $\left|\omega+2\Delta\omega_3\right|>\left|\omega-2\Delta\omega_3\right|$,
which implies that the LHS curve intersects the RHS curve first at $\theta=\pi$, and then at $\theta=0$.
For larger $m$ only the leftmost intersection remains with the two intersections with $\theta>\pi/2$
disappearing simultaneously. This situation is repeated for the second equation \eqref{eqnnlr105},
for which the RHS at $\theta=0,\pi$ gives $-I_3\left[\omega-2(\Delta\omega_2-\Delta\omega_3)\right]$
and $-I_3\left[\omega+2(\Delta\omega_2-\Delta\omega_3)\right]$ respectively; for the
fields employed we have for both molecules that
$\left|\omega-2\left(\Delta\omega_2-\Delta\omega_3\right)\right|>\left|\omega
+2\left(\delta\omega_2-\Delta\omega_3\right)\right|$.

For pyridazine the situation is different. From Fig.\ \ref{fignlr29}(a) it is clear that
the RHS curve of \eqref{eqnnlr105.1} is intersected first at $\theta=\pi$,
but the two solutions with $\theta>\pi/2$ will now disappear before the dashed curve intersects
the solid curve at $\theta=0$. For \eqref{eqnnlr105.2}
there is only one solution, which moves from $\theta=0$ to larger values of $\theta$ as the
value of $m$ increases.

Figures \ref{fignlr28.0} to \ref{fignlr29} explain the observed bifurcation structure
of the solutions of equations \eqref{eqnnlr105}. The two panels in Figures \ref{fignlr28.0}
and \ref{fignlr28} and panel \ref{fignlr29}(a) show the existence of a nonbifurcating solution
and a pair of solutions emerging from a saddle-node bifurcation.
A global view of the bifurcations is obtained rearranging equations \eqref{eqnnlr105} to
\begin{subequations}\label{eqnnlr106}
\begin{align}
\cos\theta\left(1-i_1\right)m^2+I_3(\omega+2\Delta\omega_3\cos\theta)\left[i_1+\left(1-i_1\right)\cos^2\theta\right]^2&=0,\label{eqnnlr106.1}\\
\cos\theta\left(1-i_2\right)m^2+I_3\left[\omega-2(\Delta\omega_2-\Delta\omega_3)\cos\theta\right]\left[i_2+\left(1-i_2\right)\cos^2\theta\right]^2&=0,\label{eqnnlr106.2}
\end{align}
\end{subequations}
and then plotting the zero contours of the LHS in the $\theta$-$m$ space.
These contours are shown in Fig.\ \ref{fignlr30} for the molecules treated here.
The contours indicate that a solution branch for both equations \eqref{eqnnlr106}
begins at $\theta=0$, branching out of the $\mathcal{A}_1$ solution.
On the other side, the right branch from the saddle-node pair begins at $\theta=\pi$,
which means that this solution emerges from the $\mathcal{A}_2$ solution.
The left branch of the saddle-node pair is connected only to its right partner.

The last solution set is obtained after removing all common factors in equations \eqref{eqnnlr101},
and rearranging to obtain $m^2$ as function of $\theta$ and $\psi$,
\begin{subequations}\label{eqnnlr107}
\begin{align}
m^2&=\frac{2(\Delta\omega_2\sin^2\psi-\Delta\omega_3)\cos\theta-\omega}{\cos\theta\left(I_3-I_1\cos^2\psi-I_2\sin^2\psi\right)}[({\mathsf I}\tilde{\bm C}_3)
\cdot\tilde{\bm C}_3]^2,\label{eqnnlr107.1}\\
m^2&=\frac{2\Delta\omega_2}{I_1-I_2}[({\mathsf I}\tilde{\bm C}_3)\cdot\tilde{\bm C}_3]^2,\label{eqnnlr107.2}
\end{align}
\end{subequations}
with $({\mathsf I}\tilde{\bm C}_3)\cdot\tilde{\bm C}_3$ given by equation \eqref{eqnnlr102}.
The solution is obtained by finding all the values of $\theta$ and $\psi$ such that
\begin{equation}\label{eqnnlr108}
\frac{2(\Delta\omega_2\sin^2\psi-\Delta\omega_3)\cos\theta-\omega}{\cos\theta\left(I_3-I_1\cos^2\psi-I_2\sin^2\psi\right)}=\frac{2\Delta\omega_2}{I_1-I_2},
\end{equation}
which can be rearranged and simplified to get
\begin{equation}\label{eqnnlr109}
2\cos\theta\left(\frac{I_1-I_3}{I_1-I_2}-\frac{\Delta\omega_3}{\Delta\omega_2}\right)=\frac{\omega}{\Delta\omega_2}.
\end{equation}
In terms of the polarizability this gives
\begin{equation}\label{eqnnlr110}
\cos\theta=\frac{\omega}{2\Delta\omega_2}\left(\frac{I_1-I_3}{I_1-I_2}-\frac{\alpha_3-\alpha_1}{\alpha_2-\alpha_1}\right)^{-1}.
\end{equation}
In contrast to equation \eqref{eqnnlr97}, equation \eqref{eqnnlr110} depends on the fields.
For the fields used in iodobenzene we obtain $\theta=0.4896\pi$, while for pyridazine there is no solution.

The range of $m$ for this solution is obtained from equation \eqref{eqnnlr107.2}.
The smallest value of $m$ occurs at $\psi=\pi/2,3\pi/2$, the largest at $\psi=0,\pi$,
giving the values
\begin{equation}\label{eqnnlr111}
M_i=\left(\frac{2\Delta\omega_2}{I_1-I_2}\right)^{1/2}\left[I_i+\frac{\omega^2(I_3-I_i)}{4\Delta\omega_2^2}\left(\frac{I_1-I_3}{I_1-I_2}-\frac{\Delta\omega_3}
{\Delta\omega_2}\right)^{-2} \right],
\end{equation}
where $i=1$ for $\psi=0,\pi$, and $i=2$ for $\psi=\pi/2,3\pi/2$.
A solution of \eqref{eqnnlr107} therefore exists for $m\in[M_2,M_1]$.
Noting equations \eqref{eqnnlr105}, and keeping in mind equation \eqref{eqnnlr108},
we conclude that this solution coincides with the solution of \eqref{eqnnlr107.1} at $m=M_2$,
and with the solution of \eqref{eqnnlr107.2} at $m=M_1$.

Using equations \eqref{eqnnlr107.2} and \eqref{eqnnlr111} we obtain
\begin{equation}\label{eqnnlr112}
\sin^2\theta\sin^2\psi=\frac{M_1-m}{I_1-I_2}\left(\frac{I_1-I_2}{2\Delta\omega_2}\right)^{1/2}.
\end{equation}
This equation, together with \eqref{eqnnlr109} and \eqref{eqnnlr107.2} can be substituted into
the effective potential \eqref{eqnnlr68} to get the $V_m$-$m$ curve
\begin{equation}\label{eqnnlr113}
V_m=(m-M_1/2)\left(\frac{2\Delta\omega_2}{I_1-I_2}\right)^{1/2}-\omega\cos\theta-\Delta\omega_3\cos^2\theta,
\end{equation}
where the last two terms are to be evaluated using equation \eqref{eqnnlr110}.
The first term shows a linear dependence on $m$.
In fact, as seen above, this solution is a bridge between the solutions of \eqref{eqnnlr105}
at $\psi=\pi/2,3\pi/2$ and $\psi=0,\pi$. Although it has not been considered, for negative values of $m$
the solutions are symmetric with respect to reflection on the $m=0$ line.

The $E$-$m$ diagram for iodopentafluorobenzene is shown in Fig.\ \ref{fignlr31.0}.
There are 16 different regions delimited by the $V_m$-$m$ curves.
In the figure there are two parabolas corresponding to the solution sets $\mathcal{A}_1$ (red) and 
$\mathcal{A}_2$ (green), the $\mathcal{A}_1$ parabola is always below the $\mathcal{A}_2$ as
should be evident from equations \eqref{eqnnlr104}. Emerging from the
$\mathcal{A}_1$ parabola at $m \approx 10$ there are two curves, one blue the other magenta.
These curves correspond to the independent branch solutions of equation \eqref{eqnnlr106},
the blue curve is the solution for $\psi=0,\pi$, the magenta one for $\psi=\pi/2,3\pi/2$.
These two curves intersect near $m\approx 50$, and near this intersection they are connected by
the bridge solution \eqref{eqnnlr113} (cyan). Emerging from the $\mathcal{A}_2$ curve,
near $m\approx 10$, there are also two curves with the same colors as above;
these are the two saddle-node bifurcation solutions of
equation \eqref{eqnnlr106} with the blue curve for the $\psi=0,\pi$ solution and the
magenta curve for the $\psi=\pi/2,3\pi/2$ solution.
The saddle-node bifurcation are observed as cusps near $m\approx40$ for the
blue curve and $m\approx25$ for the cyan curve. The complete $E$-$m$ diagram is
obtained when negative values of $m$ are considered, the diagram is symmetric about the $m=0$ axis.
For the full diagram the bifurcation partners form ``smiles'' that are typical features of this type
of system \cite{Tatarinov74} (for analogous structures in the case of diatomic molecules in
collinear fields, see \cite{Arango04}).

The accessible $\theta$-$\psi$ configuration space for the different regions of Fig.\ \ref{fignlr31.0}
is shown in Fig.\ \ref{fignlr32}.
Comparing with the figure for the laser interaction only (Fig.\ \ref{fignlr26}),
we see that the genus \emph{2,2}, panel \ref{fignlr26}(d), is not present when the electric field is turned on;
instead, there is a genus \emph{1,2} region, panel \ref{fignlr32}(l).
This indicates that the effect of the electric field is relatively strong compared to the laser
and that it tends to align the molecule dipole along the space fixed $z$-axis.
This is also the case for the genus \emph{3} region of Fig.\ \ref{fignlr32}(j),
which is absent in Figure \ref{fignlr26}. Note also that the genus \emph{1,1,1,1}
configurations are different in both figures.

For iodobenzene the $E$-$m$ diagram, Fig.\ \ref{fignlr31}, is simpler than the
one for iodopenfluorobenzene, Fig.\ \ref{fignlr31.0}. Since the rotational constants for
this molecule are considerably larger than those for iodopentafluorobenzene,
the $E$-$m$ diagram extends over a smaller energy range.
There are only few qualitative changes with respect to Fig.\ \ref{fignlr31.0}.
The most important difference is that the lower smile lies completely within
the \emph{2} region and does not go over to the \emph{1,1} region as for iodopentafluorobenzene.
This eliminates the lower \emph{2} region observed in the lower panel of Fig.\ \ref{fignlr31.0} and
also the \emph{1,1} region in the same panel.

The $E$-$m$ diagrams for pyridazine are shown in Fig.\ \ref{fignlr33}.
In panel (a), where  $\omega/B_3=20$, $\Delta\omega_2/B_3=22.56$, and $\Delta\omega_3/B_3=22.88$,
we obtain a very simple $E$-$m$ diagram with only 6 regions.
There is only one ``smile'' with genera \emph{2} and \emph{3}. The second smile does not form since
there is only one branch for the $\psi=\pi/2,3\pi/2$ solution. In panel (b) we show the $E$-$m$ diagram
for the same value of electric field but for a  laser field 10 times more intense. In this case,
the smile has grown bigger, overlapping the regions with genera \emph{1}, \emph{2}, and \emph{1,1}.
It is observed that below and above the overlapped regions the genus is the same as for the
weaker laser field case, while inside the smile the genus can vary according to the $E$-$m$ values.

\newpage

\section{Summary and conclusions}
\label{conclusions}

In this paper we have studied the classical mechanics of rotational motion of
dipolar asymmetric top molecules in strong external fields.  Static electric fields, linearly polarized
nonresonant laser fields, and collinear combinations of the two were investigated.
The particular asymmetric top molecules iodobenzene, pyridazine
and iodopentafluorobenzene have been treated
for physically relevant field strengths.

Following Katok \cite{Katok72} and Tatarinov \cite{Tatarinov74}, we have computed diagrams of relative equlibria in the $E$-$m$ plane; the
relative equilibria correspond physically to periodic motions with the two
Euler angles $\theta$ and $\psi$ constant \cite{Katok72,Arnold88}.
We have also examined the classically allowed $\theta$-$\psi$ configuration space for different regions of
the $E$-$m$ diagrams, and have classified the configuration space topology according to their
genus \cite{Arnold88}.  We anticipate that this classical mechanical investigation will be
useful in the difficult problem  of assigning quantum mechanical eigenstates and
energy levels for asymmetric tops in external fields \cite{Moore99,Kanya04}.

\acknowledgments

We are grateful to Mr.\ Michael Zukovsky for providing partial translations of references
\cite{Katok72} and \cite{Tatarinov74}.


\newpage

\begin{center}
\begin{table}[htbp]
\caption{\label{tabnlr1}Molecular parameters.}
\begin{center}
\begin{tabular}{ccccccccc}
\hline
Molecule    & $B_3$\footnote{Rotational contant in cm$^{-1}$}   &  $B_1/B_3$     & $B_2/B_3$   &
$d_0$\footnote{Dipole moment in Debye}
& $\alpha_1$\footnote{Polarizabilities in {\AA}$^3$} & $\alpha_2$  & $\alpha_3$ & Point group \\
\hline
C$_6$H$_5$I\footnote{Iodobenzene \cite{Poulsen04,Bulthuis97}}  & 0.189  & 0.117     
& 0.132    & 1.70    & 10.2       & 15.3        & 21.5   & $C_{2v}$\\
C$_4$H$_4$N$_2$\footnote{Pyridazine \cite{Innes88,Hinchliffe94}}  & 0.208  & 0.490     
& 0.970    & 4.14    & 5.84       & 10.29       & 10.35  & $C_{2v}$\\
C$_6$F$_5$I\footnote{Iodopentafluorobenzene \cite{Poulsen04}}   & 0.034  & 0.264   
& 0.359    & 1.54\footnote{{\it{ab-initio}} 3-21G}    & 10.5       & 17.9        & 23.8   & $C_{2v}
$\\
\hline
\end{tabular}
\end{center}
\end{table}
\end{center}

\newpage

\newpage

\begin{center}
\begin{table}[htbp]
\caption{\label{tabnlr2}Hamiltonian parameters and asymmetries with respect to $y$-axis.}
\begin{center}
\begin{tabular}{lcccccc}
\hline
Molecule    & $\omega/B_3$   &  $\Delta\omega_2/B_3$     & $\Delta\omega_3/B_3$ & $\frac{2(\Delta\omega_2-\Delta\omega_3)}{B_3}$ & $\frac{\alpha_3-\alpha_2}
{\alpha_1-\alpha_2}$ & $\frac{I_3-I_2}{I_1-I_2} $ \\
\hline
C$_6$H$_5$I & 10  & 28.46      & 63.05  & -69.88 & -1.22  &  -6.61 \\
C$_4$H$_4$N$_2$ & 20  & 22.56      & 22.88 & -0.64 & -0.014&  -0.045 \\
C$_6$F$_5$I\  & 15  & 228.0   & 409.8 & -363.58 & -0.797 & -1.785\\
\hline
\end{tabular}
\end{center}
\end{table}
\end{center}

\newpage

%
%

\newpage

\section*{Figure captions}

\begin{figure}[H]
\caption{\label{fignlr19}$E$-$m$ diagrams for free asymmetric top molecules (Euler tops).
(a) Iodobenzene; (b) Pyridazine; (c) Iodopentafluorobenzene.}
\end{figure}

\begin{figure}[H]
\caption{\label{fignlr20}Regions of classically allowed motion in
$\theta$-$\psi$ configuration space for iodopentafluorobenzene for the different
regions of Figure \ref{fignlr19}(c).}
\end{figure}

\begin{figure}[H]
\caption{\label{fignlr20a}Definition of body-fixed coordinate frames
for iodobenzene (upper panel), pyridazine (middle panel), and iodopentafluorobenzene (lower panel).
Hydrogen and fluorine atoms are represented by white circles, carbon atoms by black circles, and
nitrogen and iodine atoms by grey circles, respectively.
In each case the $x$-axis is perpendicular to the plane of the molecule, and
the dipole moment points along the body-fixed $z$-axis.}
\end{figure}

\begin{figure}[H]
\caption{\label{fignlr21}Plots of RHS (dashed), and LHS (solid) of
equations \eqref{eqnnlr80} for $m$ values at which curves do not intersect.
Left panel: equation \eqref{eqnnlr80.1}; right panel: equation \eqref{eqnnlr80.2}.}
\end{figure}

\begin{figure}[H]
\caption{\label{fignlr22}$E$-$m$ and $\theta$-$m$ bifurcation diagrams for
dipolar asymmetric top molecules in a static electric field.
(a), (b) iodobenzene ($\omega/B_3=10$); (c), (d) pyridazine ($\omega/B_3=20$);
(e), (f) iodopentafluorobenzene ($\omega/B_3=20$).}
\end{figure}

\begin{figure}[H]
\caption{\label{fignlr23}Plots of RHS (dashed), and LHS (solid) of equations \eqref{eqnnlr88}
for $m$ values at which curves do not intersect. Left panel: equation \eqref{eqnnlr88.1}
($i_1 = 1.25$, $(1-i_1)m^2 = -5$, $2 \Delta\omega_3 I_3 = 10$);
right panel: equation \eqref{eqnnlr88.2} ($i_2 = 1.2$, $(1-i_2)m^2 = -4$,
$2(\Delta\omega_2-\Delta\omega_3)I_3 = 7$).}
\end{figure}

\begin{figure}[H]
\caption{\label{fignlr24.0}$E$-$m$ diagram for the asymmetric top molecule
iodopentafluorobenzene in a nonresonant laser field: $\Delta\omega_2/B_3=228.006$
and $\Delta\omega_3/B_3=409.794$.}
\end{figure}

\begin{figure}[H]
\caption{\label{fignlr24}$E$-$m$ diagram for the asymmetric top molecule iodobenzene
in a nonresonant laser field: $\Delta\omega_2/B_3=28.456$ and $\Delta\omega_3/B_3=63.05$.}
\end{figure}

\begin{figure}[H]
\caption{\label{fignlr25}$E$-$m$ diagram for the asymmetric top molecule pyridazine
in a nonresonant laser field: $\Delta\omega_2/B_3=22.56$ and $\Delta\omega_3/B_3=22.88$.}
\end{figure}

\begin{figure}[H]
\caption{\label{fignlr26}Classically allowed (red) and forbidden (green) 
$\theta$-$\psi$ configuration space and associated genera for the various
regions of Figures \ref{fignlr24.0} and \ref{fignlr24}.}
\end{figure}

\begin{figure}[H]
\caption{\label{fignlr28.0} RHS (solid line) and LHS for $m=20$ (dashed line) of equations \eqref{eqnnlr105}
for iodopentafluorobenzene.
(a) Equation \eqref{eqnnlr105.1}; (b) equation \eqref{eqnnlr105.2}. Dotted line: $m=0$.}
\end{figure}

\begin{figure}[H]
\caption{\label{fignlr28} RHS (solid line) and LHS for $m=10$ (dashed line) of equations \eqref{eqnnlr105}
for iodobenzene.
(a) Equation \eqref{eqnnlr105.1}; (b) equation \eqref{eqnnlr105.2}. Dotted line: $m=0$.}
\end{figure}

\begin{figure}[H]
\caption{\label{fignlr29} RHS (solid line) and LHS for $m=4$ (dashed line) of equations
\eqref{eqnnlr105} for pyridazine.
(a) Equation \eqref{eqnnlr105.1}; (b) equation \eqref{eqnnlr105.2}. Dotted line: $m=0$.}
\end{figure}

\begin{figure}[H]
\caption{\label{fignlr30} Zero contours of the LHS of equations \eqref{eqnnlr106}.
Iodobenzene: (a) eq.\ \eqref{eqnnlr106.1}, (b) eq.\ \eqref{eqnnlr106.2}.
Pyridazine: (c) eq.\ \eqref{eqnnlr106.1}; (d) eq.\ \eqref{eqnnlr106.2}.
Iodopentafluorobenzene: (e) eq.\ \eqref{eqnnlr106.1}, (f) eq.\ \eqref{eqnnlr106.2}.}
\end{figure}

\begin{figure}[H]
\caption{\label{fignlr31.0} $E$-$m$ diagram for the dipolar asymmetric top molecule
iodopentafluorobenzene in collinear static electric and nonresonant laser fields.
$\omega/B_3=20$, $\Delta\omega_2/B_3=228.006$ and $\Delta\omega_3/B_3=409.7945$.}
\end{figure}

\begin{figure}[H]
\caption{\label{fignlr32}
Classically allowed (red) and forbidden (green) $\theta$-$\psi$
configuration space and associated genera for the regions of Figure \ref{fignlr31.0}.
Genus \emph{0} not included.}
\end{figure}

\begin{figure}[H]
\caption{\label{fignlr31} $E$-$m$ diagram for the dipolar asymmetric top molecule
iodobenzene in collinear static electric and nonresonant laser fields.
$\omega/B_3=10$, $\Delta\omega_2/B_3=28.456$ and $\Delta\omega_3/B_3=63.05$.}
\end{figure}

\begin{figure}[H]
\caption{\label{fignlr33} $E$-$m$ diagram for the dipolar asymmetric top molecule
pyridazine in collinear static electric and nonresonant laser fields.
(a) $\omega/B_3=20$, $\Delta\omega_2/B_3=22.56$, and $\Delta\omega_2/B_3=22.88$;
(b) same electric field but a laser field 10 times more intense.}
\end{figure}


\newpage
\begin{figure}[H]
\centering
\includegraphics[angle=0,width=4.0in]{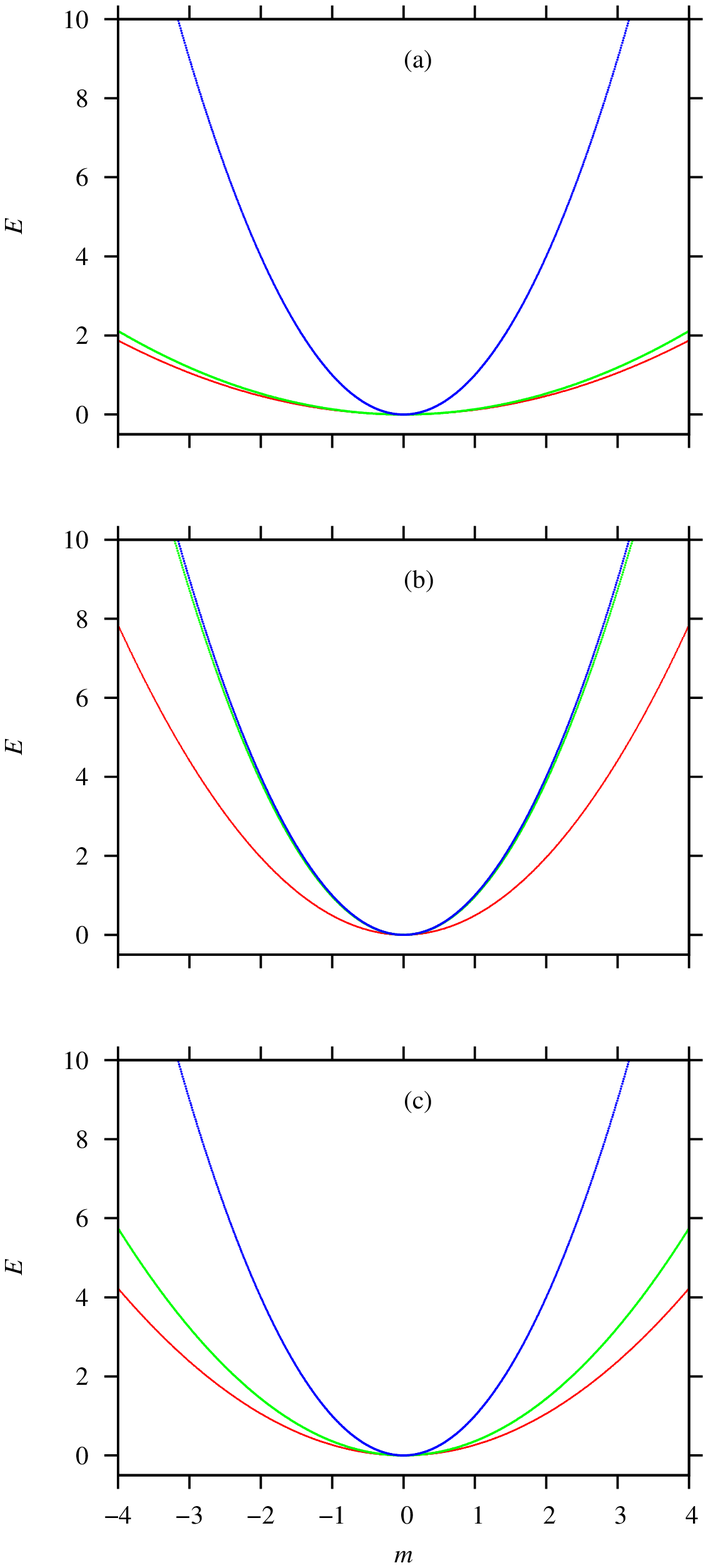}
\end{figure}

\noindent FIGURE 1  \;\;(C.\ A.\ Arango \& G.\ S.\ Ezra)

\newpage
\begin{figure}[H]
\centering
\includegraphics[width=6.5in]{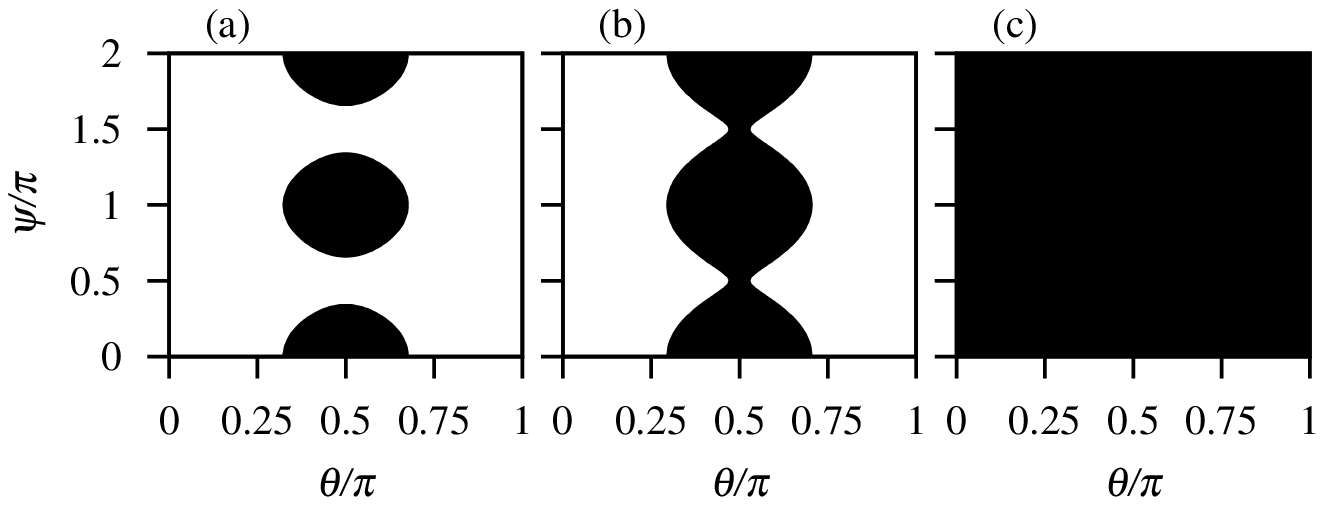}
\end{figure}

\vspace{0.5in}\noindent FIGURE 2  \;\;(C.\ A.\ Arango \& G.\ S.\ Ezra)

\newpage
\begin{figure}[H]
\centering
\includegraphics[width=4.0in]{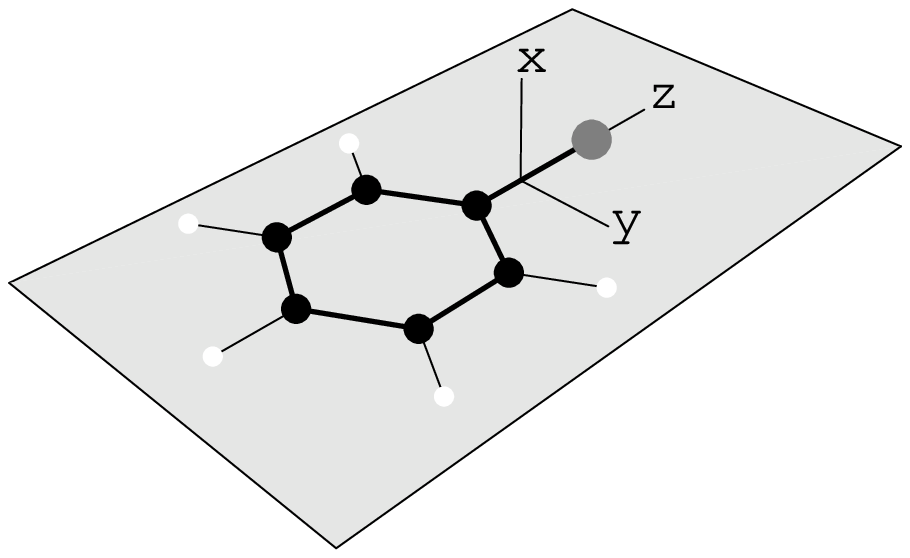}\vspace{-1.5cm}
\includegraphics[width=4.0in]{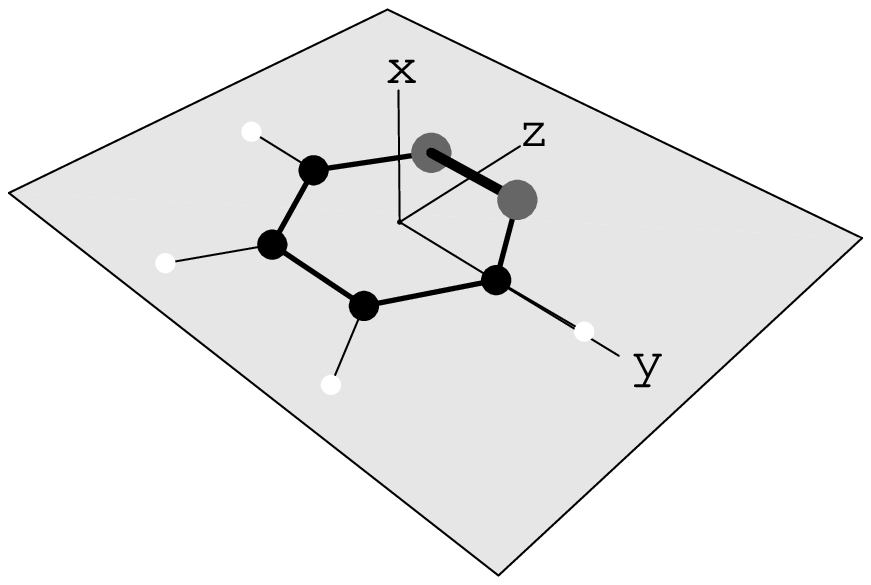}\vspace{-1cm}
\includegraphics[width=4.0in]{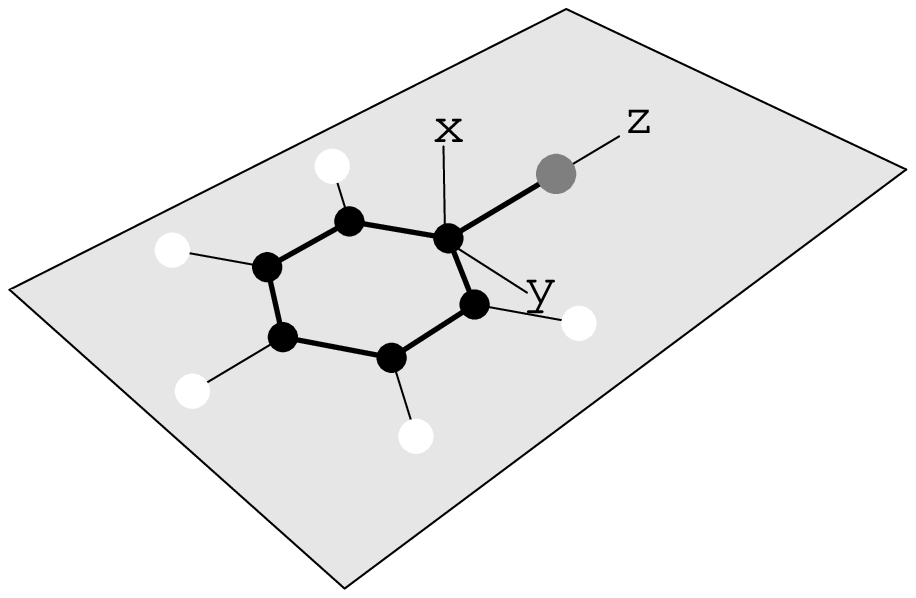}
\end{figure}

\vspace{0.5in}\noindent FIGURE 3  \;\;(C.\ A.\ Arango \& G.\ S.\ Ezra)

\newpage
\begin{figure}[H]
\centering
\includegraphics[angle=-90,width=6.5in]{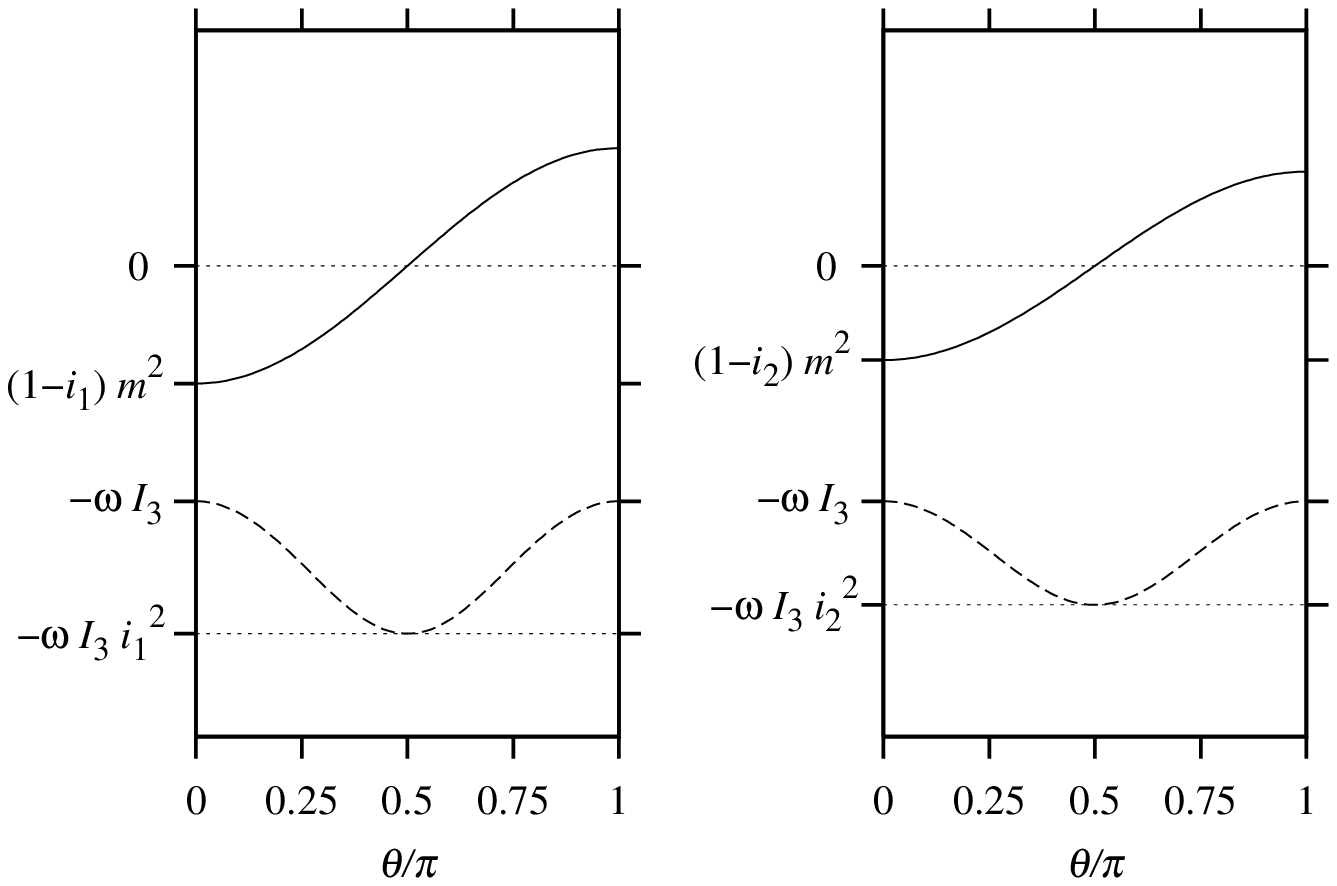}
\end{figure}

\vspace{0.5in}\noindent FIGURE 4  \;\;(C.\ A.\ Arango \& G.\ S.\ Ezra)

\newpage
\begin{figure}[H]
\centering
\includegraphics[width=6.5in]{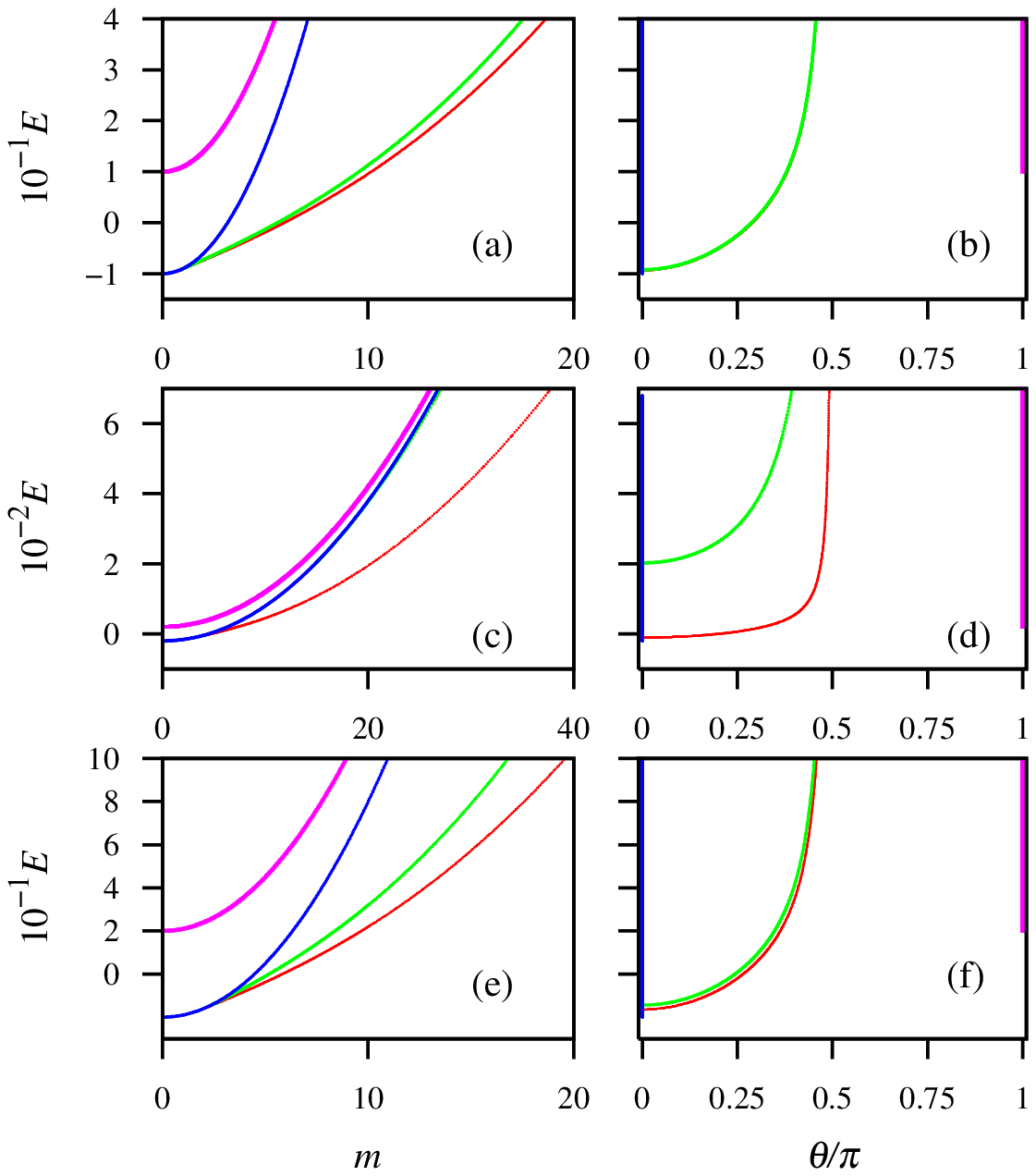}
\end{figure}

\vspace{0.5in}\noindent FIGURE 5  \;\;(C.\ A.\ Arango \& G.\ S.\ Ezra)

\newpage
\begin{figure}[H]
\centering
\includegraphics[angle=-90,width=6.5in]{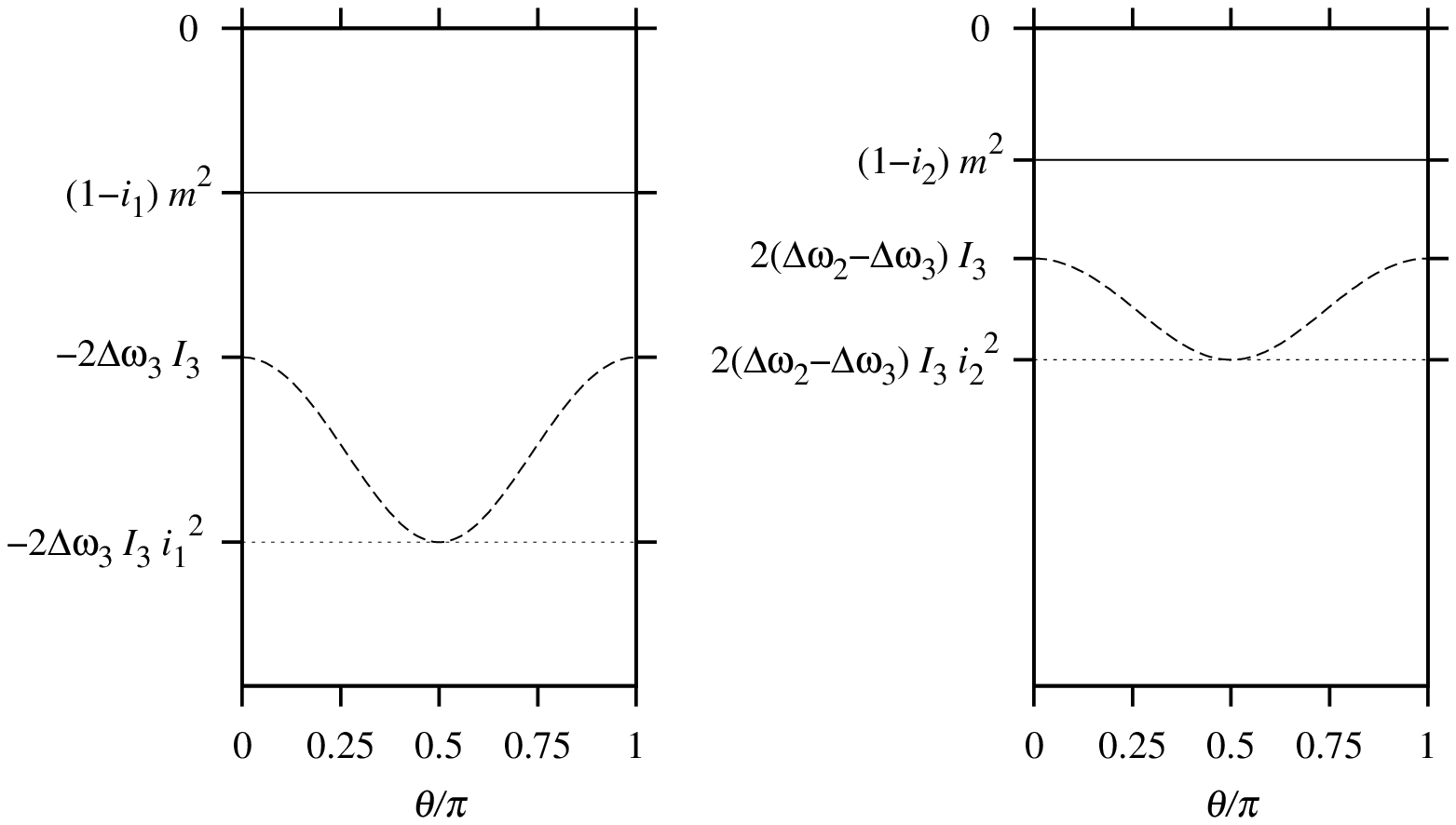}
\end{figure}

\vspace{0.5in}\noindent FIGURE 6  \;\;(C.\ A.\ Arango \& G.\ S.\ Ezra)

\newpage
\begin{figure}[H]
\centering
\includegraphics[angle=-90,width=6.5in]{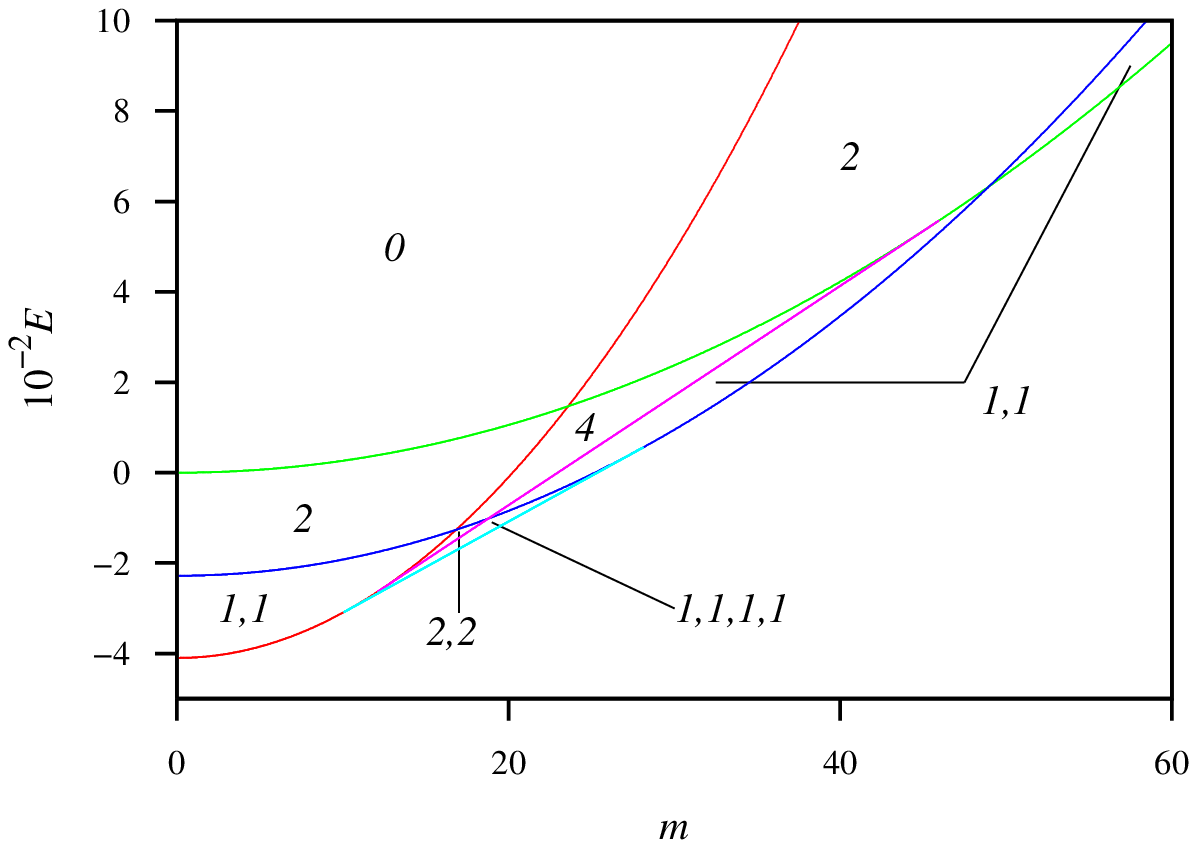}
\end{figure}

\vspace{0.5in}\noindent FIGURE 7  \;\;(C.\ A.\ Arango \& G.\ S.\ Ezra)

\newpage
\begin{figure}[H]
\centering
\includegraphics[angle=-90,width=6.5in]{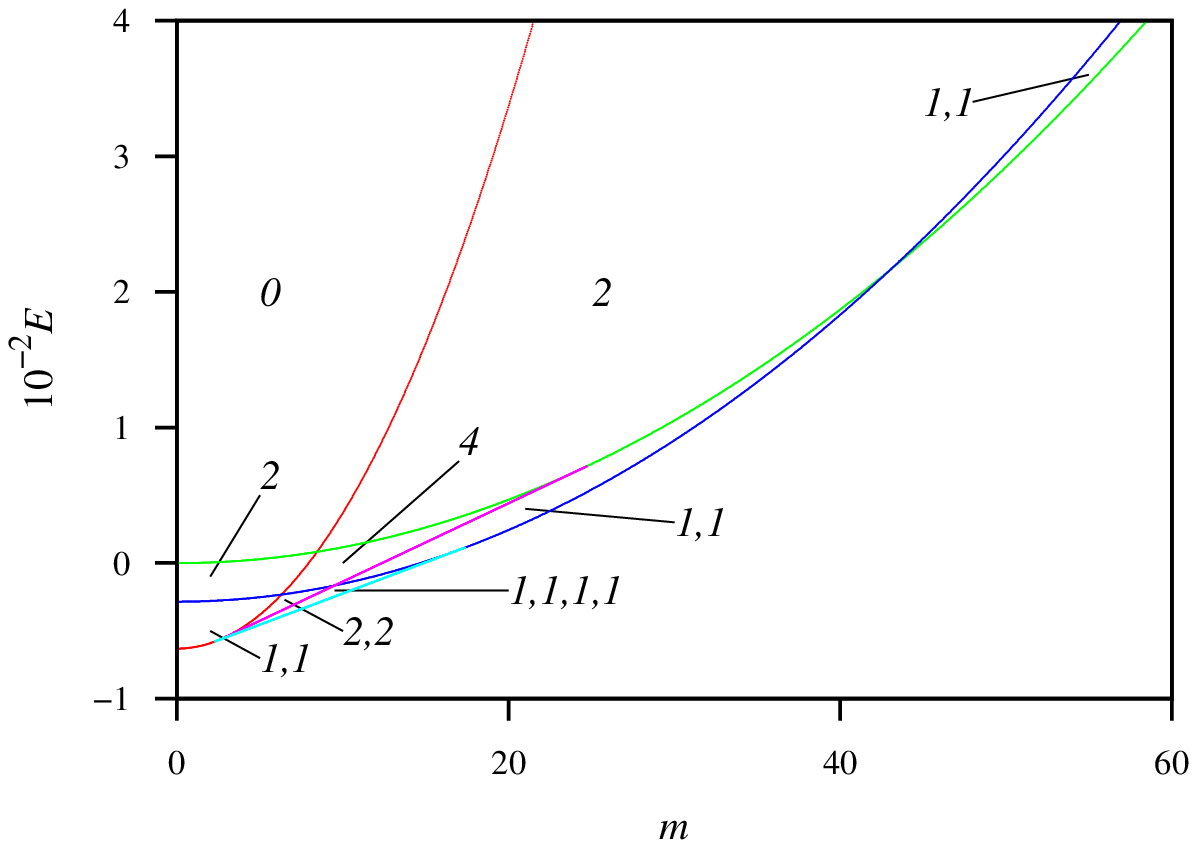}
\end{figure}

\vspace{0.5in}\noindent FIGURE 8  \;\;(C.\ A.\ Arango \& G.\ S.\ Ezra)

\newpage
\begin{figure}[H]
\centering
\includegraphics[angle=-90,width=6.5in]{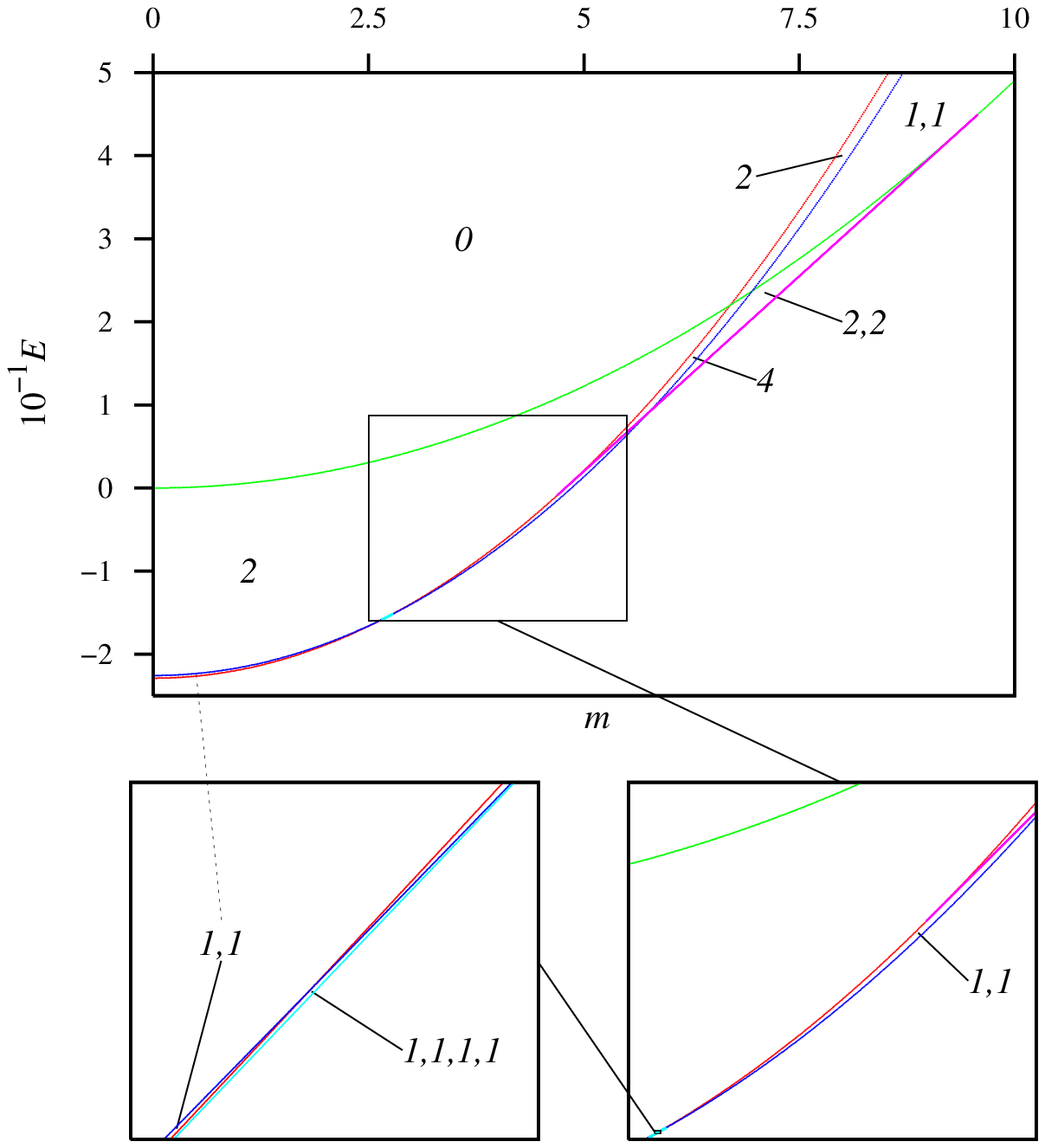}
\end{figure}

\vspace{0.5in}\noindent FIGURE 9 \;\;(C.\ A.\ Arango \& G.\ S.\ Ezra)

\newpage
\begin{figure}[H]
\centering
\includegraphics[width=6.5in]{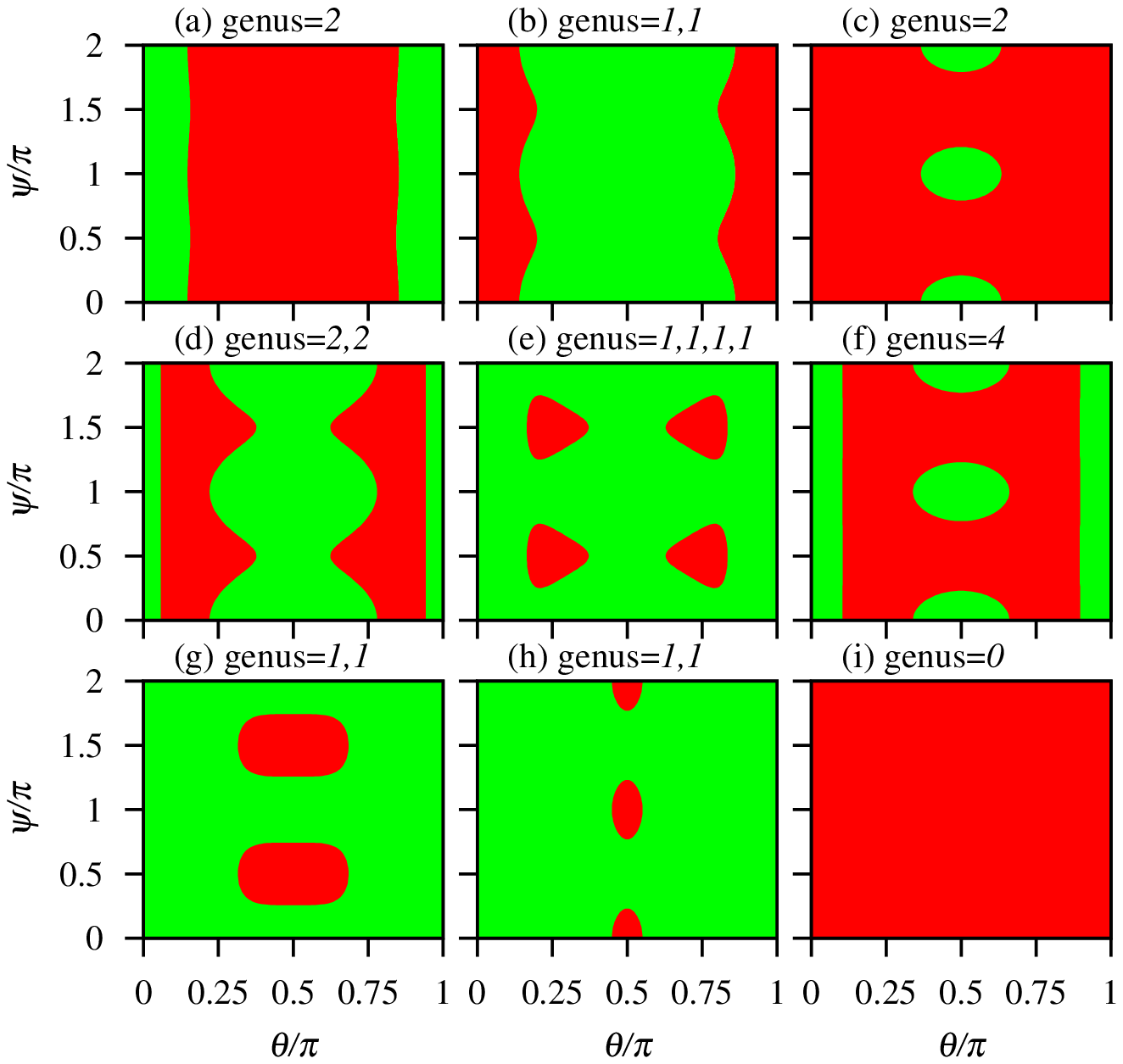}
\end{figure}

\vspace{0.5in}\noindent FIGURE 10  \;\;(C.\ A.\ Arango \& G.\ S.\ Ezra)

\newpage
\begin{figure}[H]
\centering
\includegraphics[angle=-90,width=6.0in]{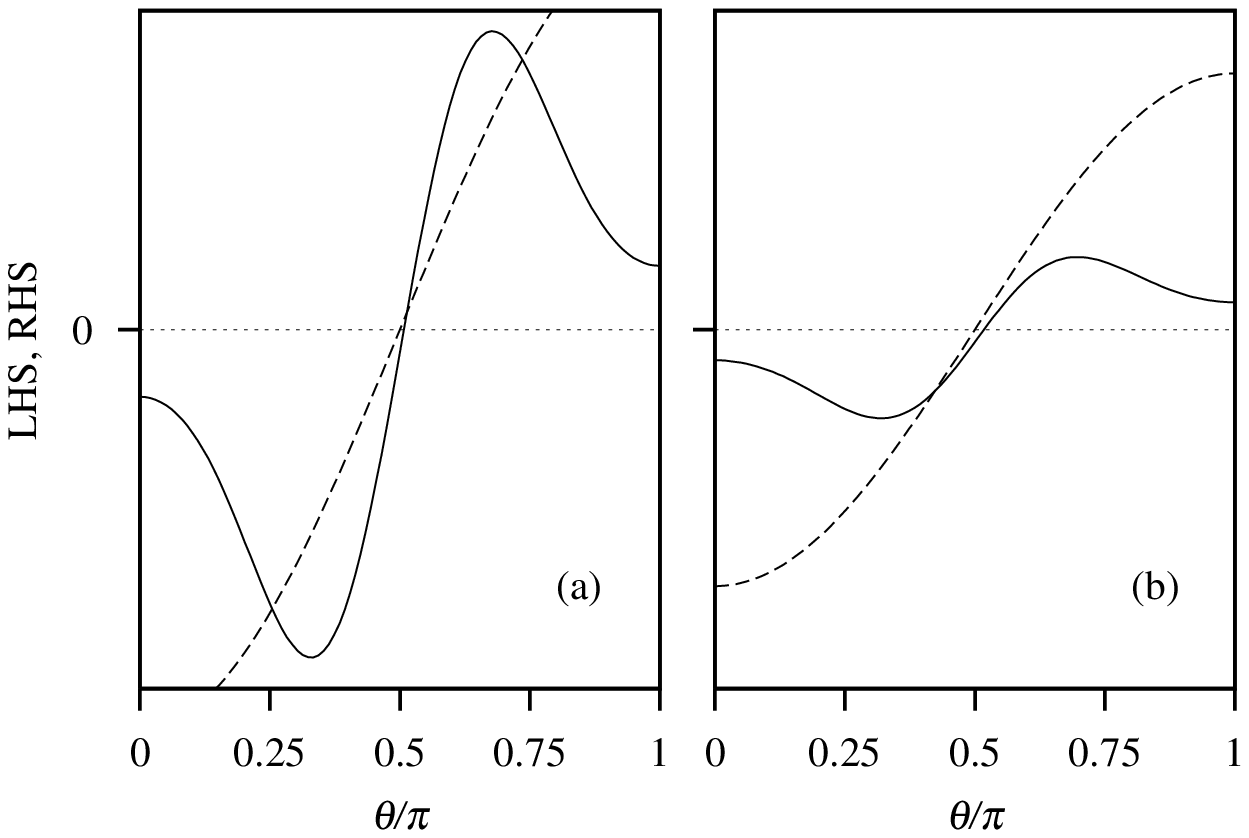}
\end{figure}

\vspace{0.5in}\noindent FIGURE 11 \;\;(C.\ A.\ Arango \& G.\ S.\ Ezra)

\newpage
\begin{figure}[H]
\centering
\includegraphics[angle=-90,width=6.0in]{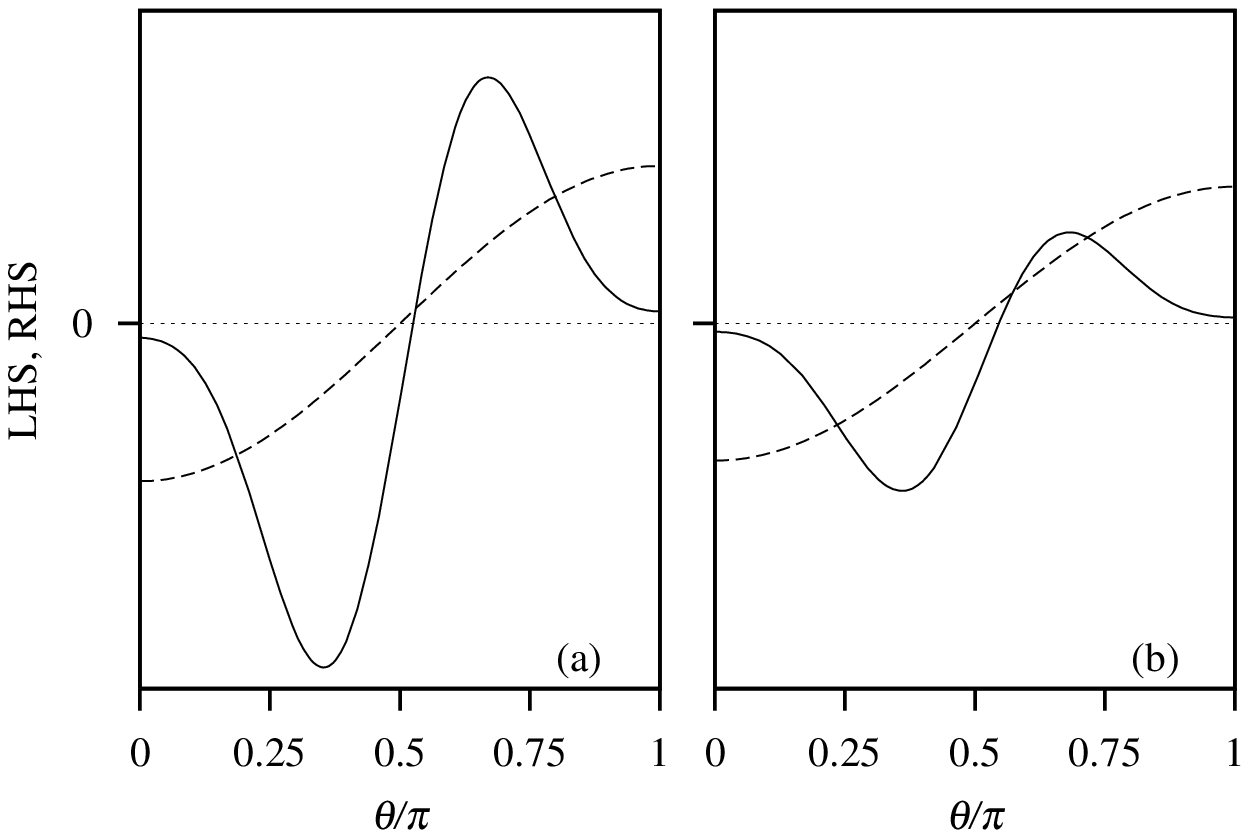}
\end{figure}

\vspace{0.5in}\noindent FIGURE 12 \;\;(C.\ A.\ Arango \& G.\ S.\ Ezra)

\newpage
\begin{figure}[H]
\centering
\includegraphics[angle=-90,width=6.0in]{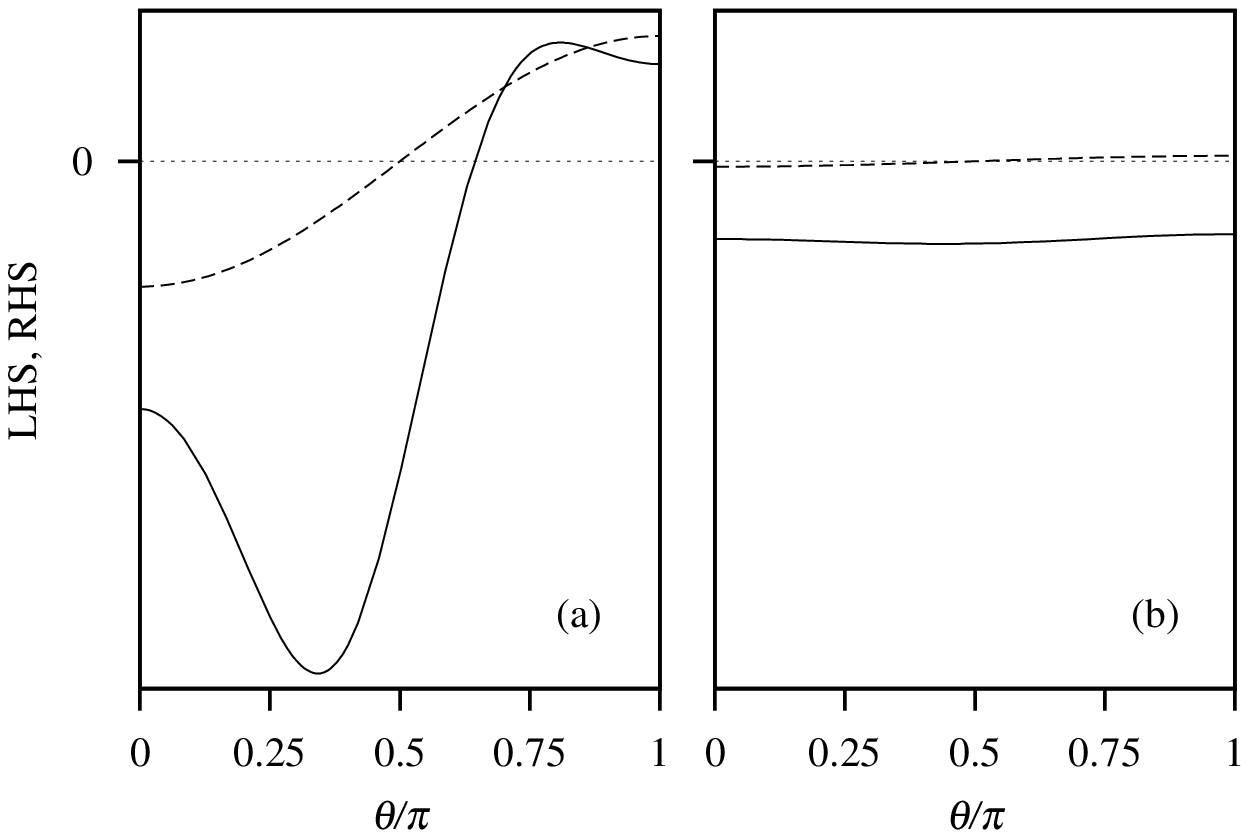}
\end{figure}

\vspace{0.5in}\noindent FIGURE 13  \;\;(C.\ A.\ Arango \& G.\ S.\ Ezra)

\newpage
\begin{figure}[H]
\centering
\includegraphics[width=6.0in]{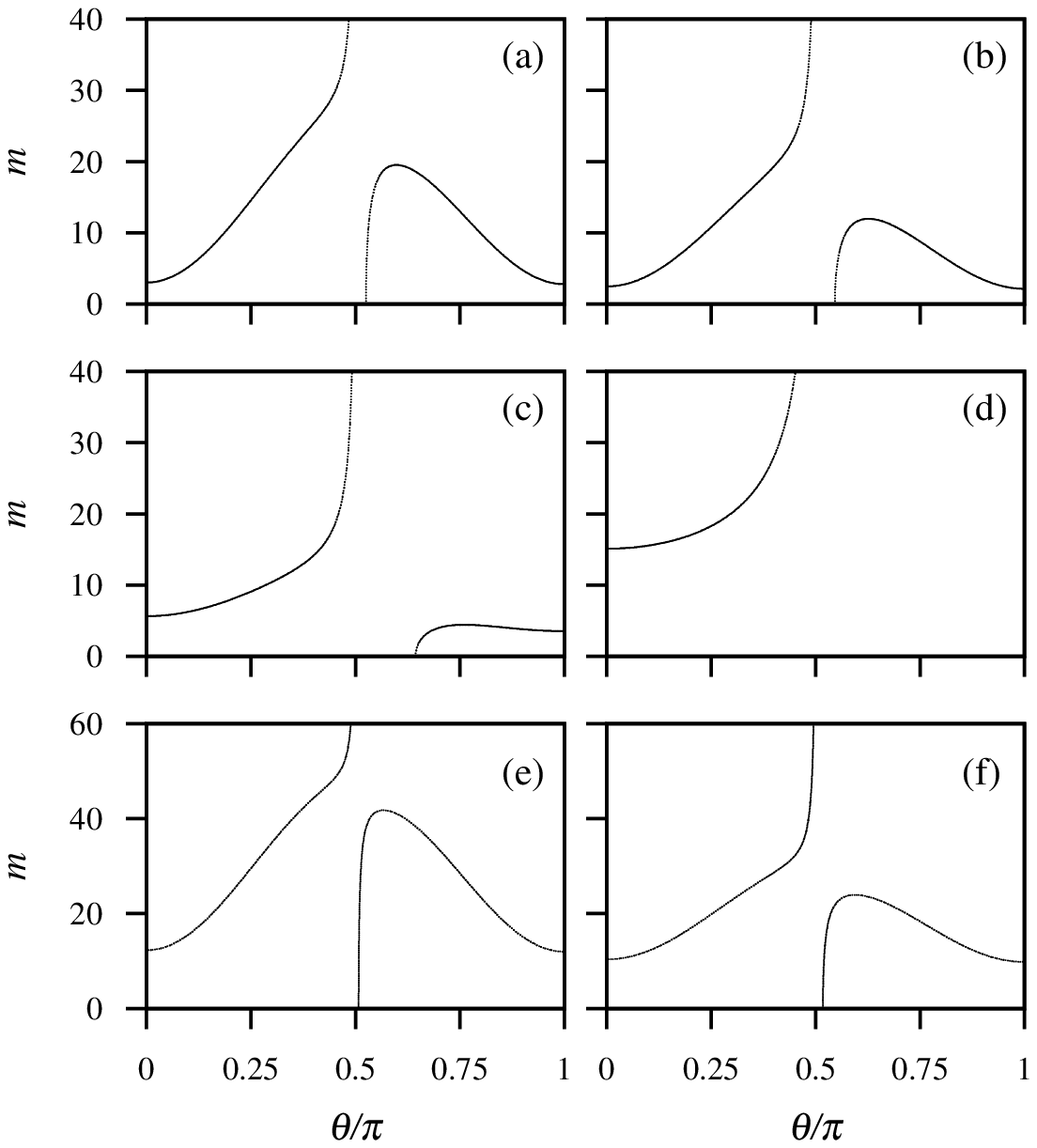}
\end{figure}

\vspace{0.5in}\noindent FIGURE 14  \;\;(C.\ A.\ Arango \& G.\ S.\ Ezra)

\newpage
\begin{figure}[H]
\centering
\includegraphics[angle=-90,width=6.0in]{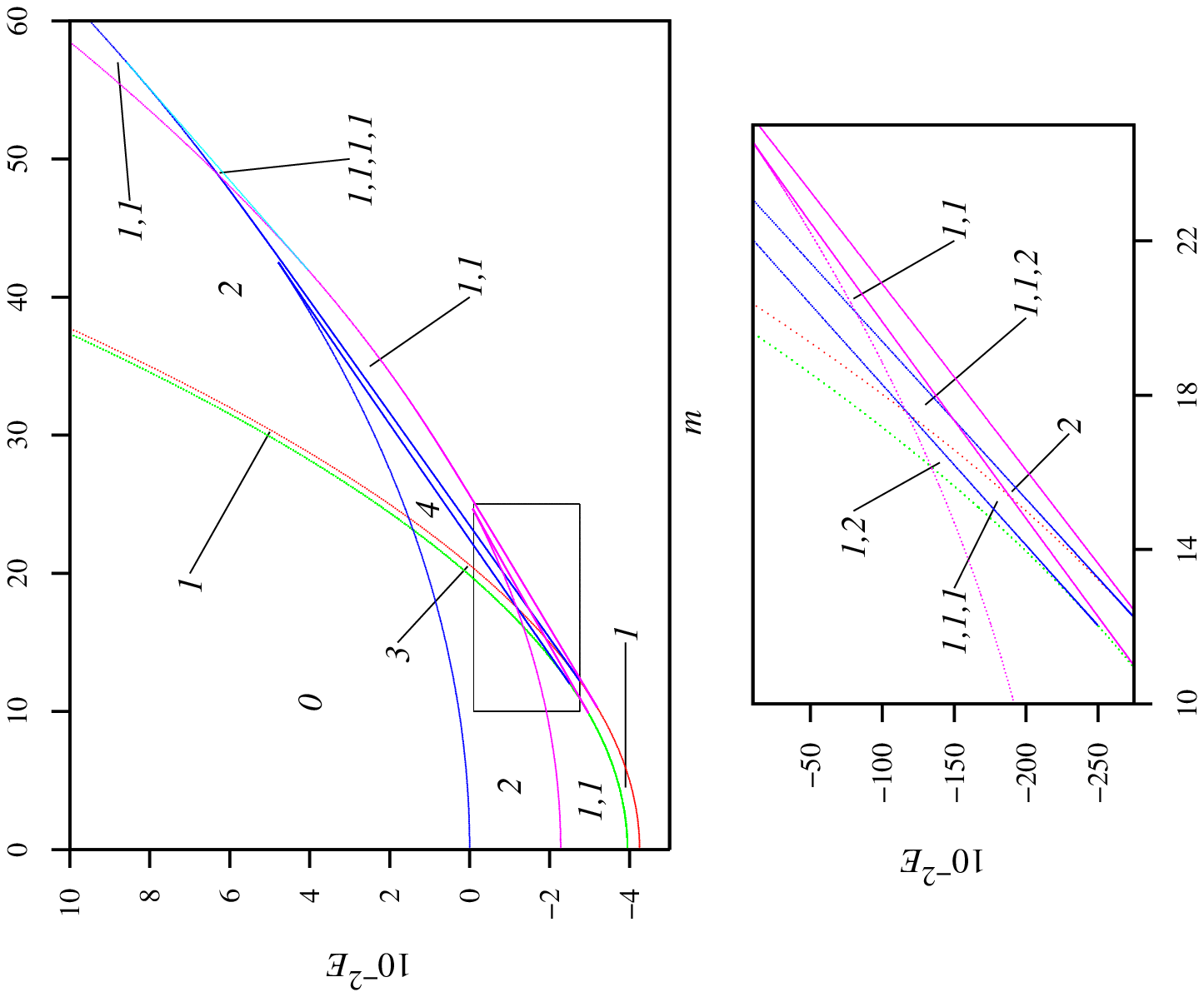}
\end{figure}

\vspace{0.5in}\noindent FIGURE 15  \;\;(C.\ A.\ Arango \& G.\ S.\ Ezra)

\newpage
\begin{figure}[H]
\centering
\includegraphics[width=6.0in]{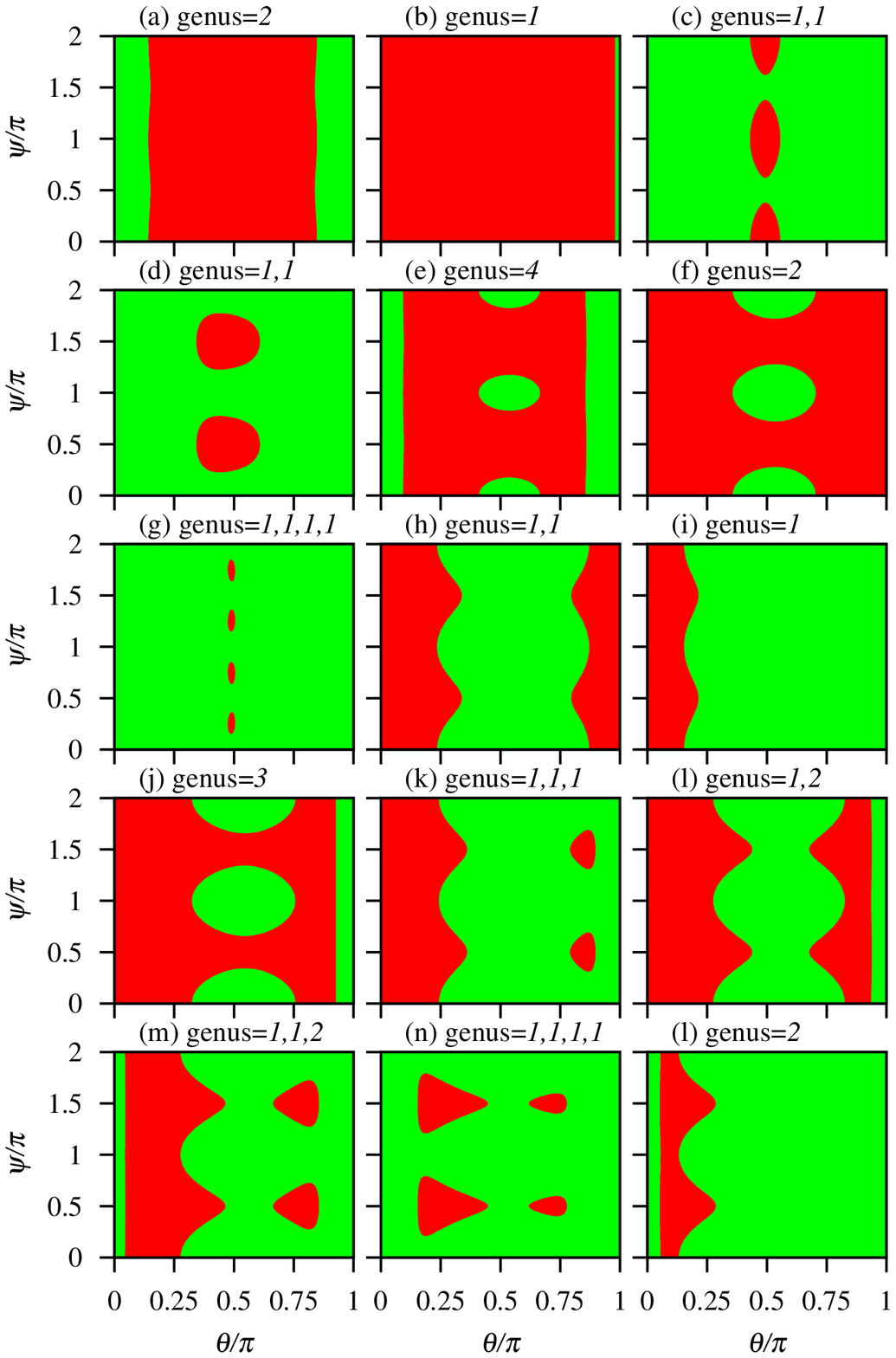}
\end{figure}

\noindent FIGURE 16  \;\;(C.\ A.\ Arango \& G.\ S.\ Ezra)

\newpage
\begin{figure}[H]
\centering
\includegraphics[angle=-90,width=6.0in]{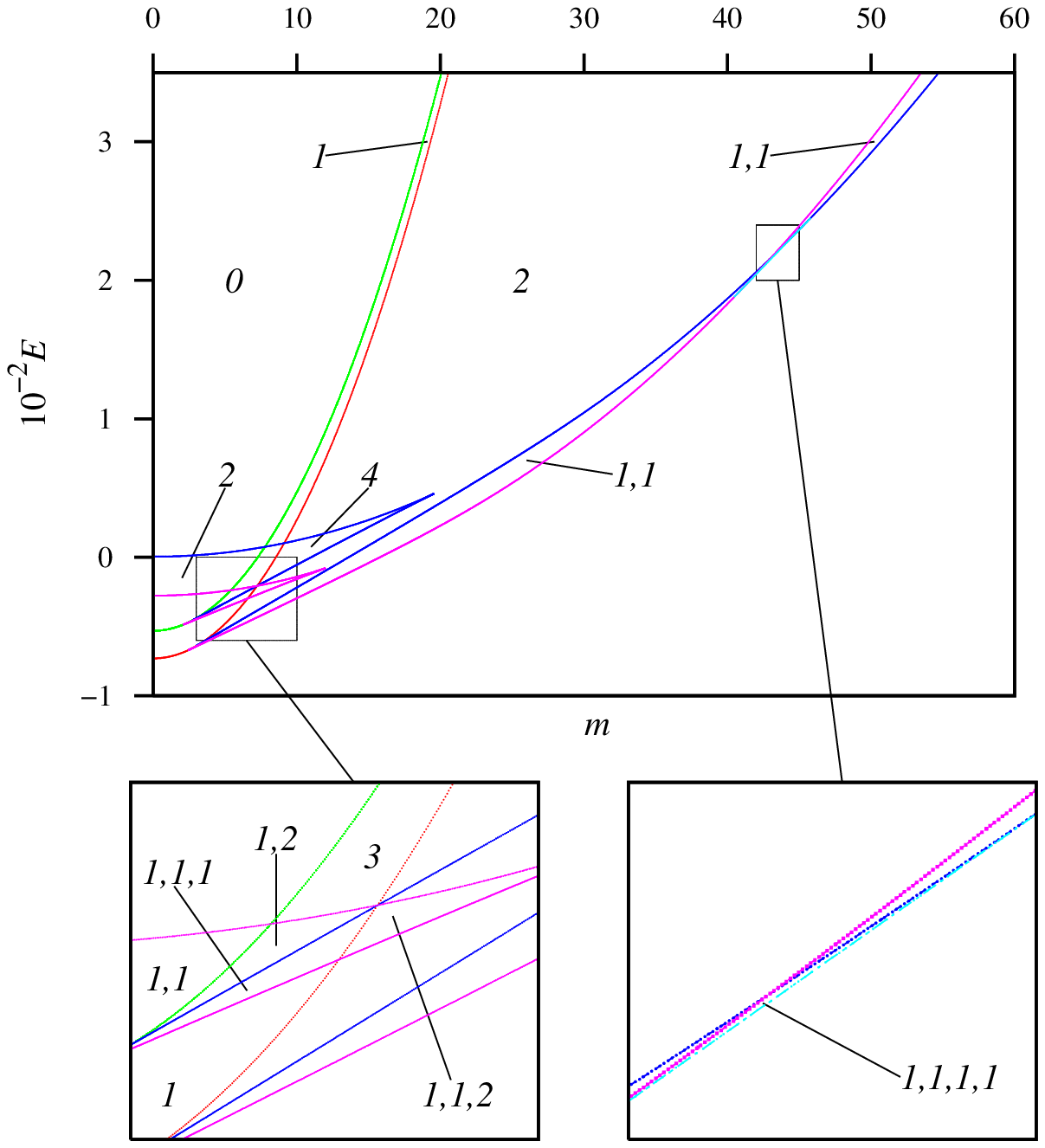}
\end{figure}

\vspace{0.5in}\noindent FIGURE 17  \;\;(C.\ A.\ Arango \& G.\ S.\ Ezra)

\newpage
\begin{figure}[H]
\centering
\includegraphics[angle=-90,width=6.0in]{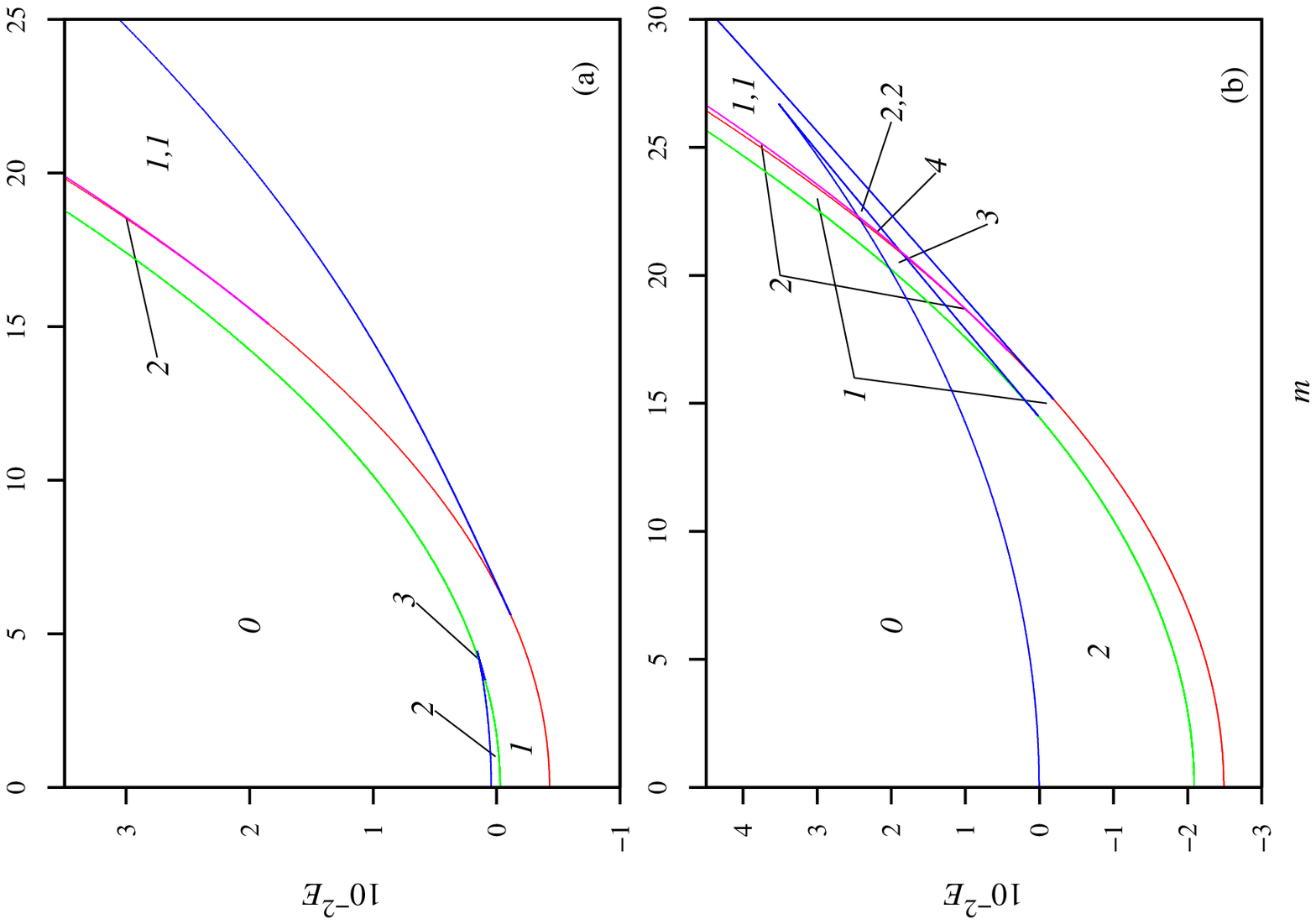}
\end{figure}

\noindent FIGURE 18  \;\;(C.\ A.\ Arango \& G.\ S.\ Ezra)

\end{document}